\documentclass[onecolumn,usenatbib]{mn2e}
\usepackage{graphicx}

\newcommand{\beq}{\begin{equation}}
\newcommand{\eeq}{\end{equation}}
\newcommand{\beqa}{\begin{eqnarray}}
\newcommand{\eeqa}{\end{eqnarray}}

\newcommand{\Tr}{{\rm Tr~}}
\newcommand{\erf}{{\rm erf~}}
\newcommand{\FWHM}{{\rm FWHM}}
\newcommand{\rmd}{{\rm d}}
\newcommand{\rme}{{\rm e}}
\newcommand{\rmi}{{\rm i}}

\title{Shear calibration biases in weak lensing surveys}

\author[{C. Hirata \& U. Seljak}]{Christopher
Hirata\thanks{E-mail: chirata@princeton.edu},
 Uro\v s Seljak\thanks{E-mail: uros@feynman.princeton.edu} \\
Department of Physics, Jadwin Hall, Princeton University, Princeton, NJ 08544, USA
}

\date{1 April 2003}

\begin{document}

\maketitle

\begin{abstract}
We investigate biases induced by 
the conversion between the observed image shape to shear
distortion in current weak lensing analysis methods. Such overall 
calibration biases cannot be detected by the standard tests
such as E/B decomposition or calibration with stars. 
We find that the non-Gaussianity of 
point spread function has a significant effect and
can lead to up to 
15 per cent error on the linear amplitude of fluctuations 
$\sigma_8$ depending on the method of analysis. This could
explain some of the discrepancies seen in recent amplitude 
determinations from weak lensing. 
Using an elliptical Laguerre expansion method we develop a 
re-Gaussianization
method which reduces the error to calibration error of
order 1 per cent even for poorly resolved galaxies. 
We also discuss a new type of shear selection bias which 
results in up to roughly 8 percent underestimation of
 the signal. It is
expected to scale with redshift, inducing errors in the 
growth factor extraction if not properly corrected for. 
Understanding and correcting for such effects
is crucial if weak lensing is to become a high precision
probe of cosmology.  

\end{abstract}


\section{Introduction}
\label{sec:intro}

Gravitational lensing is a powerful tool to probe the 
large scale matter distribution in the universe (see
e.g. \citealt{2001PhR...340..291B} for a review).  
As such it offers the
possibility of breaking the degeneracies between many 
cosmological parameters (e.g. \citealt{1999ApJ...514L..65H})
 and serves as a powerful
complementary method to the cosmic microwave background
anisotropies. Its main advantage is that it is sensitive 
to the dark matter directly and so avoids many of the 
complications present in other cosmological probes  at low 
redshift such 
as galaxy clustering or cluster surveys.
Recent first detections
of shear correlations (\citealt{2000MNRAS.318..625B},
\citealt{2000A&A...358...30V},
\citealt{2001ApJ...552L..85R},
\citealt{2002ApJ...572...55H}, \citealt{2002astro.ph.10213B},
\citealt{2002astro.ph.10604J}) have already demonstrated 
the promise of this method. Upcoming surveys such as 
NOAO deep survey ({\it www.noao.edu/noao/noaodeep/}), 
the CFHT legacy survey ({\it www.cfht.hawaii.edu/Science/CFHLS/})
and the Deep Lens Survey ({\it dls.bell-labs.com/})
will go beyond simply detecting cosmic shear but will map it over
large areas of the sky. Finally, proposed projects such 
as DMT/LSST ({\it www.dmtelescope.org/dark\_home.html}),
 Pan-STARRS ({\it www.ifa.hawaii.edu/pan-starrs/}) or {\it SNAP}
({\it snap.lbl.gov})
 will map a large fraction of the sky
with 
high depth, achieving unprecedented statistical precision. 

Advances in the statistical precision require parallel progress 
on the systematics errors if the information from the current and 
upcoming experiments is to be exploited fully. The effects one is looking 
for are extremely small, so any small errors in the calibration or 
data reduction can lead to a large error in the final result.
The most commonly used methods of reduction 
so far are based on work by \citet{1995ApJ...449..460K} (KSB95).
Recently there have been several new approaches proposed dealing 
with various aspects of this problem developed by
\citet{2000ApJ...537..555K}, \citet{2002AJ....123..583B} 
(BJ02), and \citet{2001astro.ph..5179R}. 
What is somewhat unclear from this work is how much difference these new 
approaches make and what are the remaining errors in them. 
While most of these papers present at least some tests, these are often 
done using simplified models and do not attempt to quantify the 
error as a function of size of the galaxy or their ellipticity. 
The original KSB95 method was calibrated off of simulations, and other 
authors have investigated calibration issues in some of the variants and 
extensions of KSB95 numerically (\citealt{2000ApJ...537..555K}; 
\citealt{2001A&A...366..717E}; \citealt{2001MNRAS.325.1065B}).  However, 
the calibration of more recent methodologies such as BJ02 has not yet been 
studied in detail, nor have the various methods been compared in the same 
simulation.  (Many of the simulations in the literature are not directly 
comparable since they have different assumptions about the distribution of 
galaxy shapes and sizes and the PSF and noise properties.)
The goal of this paper is to address some of these issues in more detail.
We pay particular attention to the problem of point spread function (PSF)
dilution in realistic 
PSF and galaxy shape models, which 
only affects the overall calibration of the shear (if seeing is not 
significantly variable across the lensing field) and does not show up 
in the systematics tests.

Extracting shear from the images is not the only place where a bias 
might arise in the 
weak lensing analysis. Other possibilities include (but are not limited
to) intrinsic correlations, 
errors in the source redshift distribution, and
so-called selection bias, where a galaxy is preferentially selected 
when its shape is aligned with the PSF distortion. 
However, there is
another
possible bias caused by the fact that a galaxy may also be preferentially 
selected when aligned orthogonally to the intrinsic shear, because 
detection algorithms preferentially detect circular objects. This bias 
depends on the detection threshold and the relative angular size of the 
galaxy and PSF; it becomes unimportant for poorly
resolved galaxies (i.e. galaxies much smaller than the PSF) because in 
this case the size and shape of the galaxy's image are determined by the 
PSF irrespective of the galaxy's physical orientation.  Since it is 
directly correlated 
with the signal it is very difficult to detect it using the standard
systematics
tests such as the 45 degree rotation test (E/B decomposition) 
and must be corrected for using simulations or some other method
that explicitly takes it into account. 

This paper is organized as follows. In \S\ref{sec:2} we review past
approaches
to PSF corrections, present the formalism for
PSF correction based on the elliptical Laguerre expansion, and derive two 
approximate methods to correct for non-Gaussianity of the PSF. We compare
various methods in the literature and estimate their biases 
using simulations. In \S\ref{sec:ssbsec} we discuss a shear selection bias 
and how it can affect the amplitude of fluctuations. Conclusions 
are presented in \S\ref{sec:4} and in the appendices we describe technical
details on convolution in the Laguerre expansion formalism, the linear PSF
method, and the KSB95
method.

\section{PSF Corrections}
\label{sec:2}

In actual measurements, we do not measure directly the moments of
intrinsic galaxy images, but rather we 
measure the moments of galaxy image after convolution with the PSF of the
telescope.  We will denote the intrinsic 
galaxy image by $f({\bmath x})$, the PSF by $g({\bmath x})$, and the
measured
galaxy image by $I({\bmath x})$.  Our eventual 
objective is 
to take the PSF $g({\bmath x})$ (which can be reconstructed by considering
the images of stars in the field) and the 
observed image of the galaxy $I({\bmath x})$, and construct from these the
ellipticity ${\bf e}_f$ of the intrinsic 
image that can (with an appropriate calibration factor) be used as an
estimator for the local shear. 

Many approaches to this problem appear in the literature.  The common
theme in most of them is to measure the moments of the image and PSF using
some weight function:

\beq
M_i^{(I)} = \int_{{\bf R}^2} P_i({\bmath x}) w({\bmath x}) I({\bmath
x}) \rmd^2{\bmath x},
\eeq
where ${\bmath x}=(x,y)$ is the position on the sky, $w$ is the weight
function, and the $P_i$ are polynomials in $x$ and $y$.  These moments
$M_i^{(I)}$, and their counterparts $M_i^{(g)}$ for the PSF, are then used
to compute the ellipticity estimator.  The methods mainly differ in their
answers to
the following questions:

\newcounter{Lcount01}
\begin{list}{\arabic{Lcount01}.}{\usecounter{Lcount01}}
\item\label{it:adaptive} Is the weight function a circular Gaussian, an
elliptical Gaussian, or some other shape? 
\item\label{it:psfgal} Is the weight function used for the PSF the same as
that for the galaxy, or is it matched to the size of the PSF? 
\item\label{it:deconvolve} Does the method (partially) deconvolve the
galaxy image or does it directly correct the galaxy moments for the effect
of the PSF?  In the latter
case, which corrections are performed using perturbation theory, which are
performed by comparison to simulations, and which are performed using
simple analytical results
valid in limiting cases?
\end{list}

Question \ref{it:adaptive} has typically been decided in favor of a
Gaussian weight due to its rapid convergence 
to zero at large radii (hence the polynomial moments are all
convergent)
and absence of singularities, as well as general
mathematical 
convenience.  The majority of recent theoretical
(\citealt{2000ApJ...537..555K}, \citealt{2001astro.ph..5179R}) 
and observational (\citealt{2000MNRAS.318..625B},
\citealt{2000A&A...358...30V},
\citealt{2001ApJ...552L..85R},
\citealt{2002ApJ...572...55H}) analyses
 have followed KSB95 in using a circular Gaussian weight function.  The
elliptical Gaussian weight function was introduced by 
BJ02; although it is considerably more complicated to work with, a general
ellipse can be a better match to an object's actual shape than a
circle.  This property is
important for those methods that treat the deviation of an object or PSF
from Gaussianity via perturbation theory, including our linear and
re-Gaussianization methods.

There is little consensus in the literature on the best answer to Question
\ref{it:psfgal}.  The advantage of measuring the PSF with its own weight
function, independently
of the galaxy, is that the PSF weight function can be matched to the size
(and, in the case of elliptical moments, the shape) of the PSF.  This is
necessary for methods
(e.g. re-Gaussianization) that treat the non-Gaussianity of the PSF
perturbatively.  The disadvantage of measuring moments of the PSF with a
weight function matched to the PSF is that
features of the PSF outside the scale radius of the weight function are
suppressed.  Features at radii large compared to the characteristic scale
of the PSF are present in
nearly all real-world PSFs: they are introduced as diffraction rings in
space-based observations, and as a power law ($g\propto r^{-11/3}$) tail
from Kolmogorov turbulence in ground-based observations.
If the galaxy is of size comparable to or somewhat larger than the PSF,
these features can fall inside the scale radius of the galaxy's weight
function, with potentially
disastrous results for attempts to unravel the effects of the PSF on the
measured galaxy moments (see Section \ref{sec:sim}).  The higher-order
moments of the PSF must therefore be considered in
order to take into account features at large radius if a PSF weight
matched to the size of the PSF itself is used.
Numerical tests \citep{1998NewAR..42..137H} indicate that the KSB95 method
supplemented with a pre-seeing shear polarizability calculation
\citep{1997ApJ...475...20L} works
best when the PSF is measured with the same weight function as the
galaxy; this is unsurprising, since this method does not consider any
moments higher than the fourth order (see Appendix \ref{sec:ksb}).

Finally we arrive at Question \ref{it:deconvolve}: how do we convert the
measured moments of the galaxy and PSF into an ellipticity of the
intrinsic galaxy image?  Most
methods, including \citet{1997ApJ...475...20L},
\citet{1998NewAR..42..137H}, and BJ02, have corrected the galaxy moments
for the effect of the PSF using either perturbation theory, simple
analytical
corrections, or some combination thereof.  (The original KSB95 work used
simulations to compute the diluting effect of
the PSF on the galaxy.)  \citet{2001astro.ph..5179R} developed a
deconvolution method based on the circular Laguerre expansion of the PSF
and galaxy (see Section \ref{sec:ele}), writing the PSF convolution as a
matrix acting on
the vector of galaxy moments and cutting off the hierarchy of Laguerre
moments at sixth
order to prevent noise amplification in the deconvolution (matrix
inversion).
BJ02 have investigated matrix inversion methods as well
using the raising and lowering operator formalism, with
particular attention paid to the effects of the matrix inversion
on noise.
  In this paper, we introduce and investigate two perturbative
approaches, expanding around the case of
elliptical Gaussian PSFs and galaxies.  The ``linear'' method (Appendix
\ref{sec:psfcorr}) is essentially a first-order
approximation to \citet{2001astro.ph..5179R}'s matrix inverse, generalized
to elliptical moments. It improves significantly on the same order method 
proposed by BJ02 in their Appendix C
 and should be of particular interest to those 
using Sloan Digital Sky Survey PHOTO reduction for weak lensing. 
The ``re-Gaussianization'' method (Section \ref{sec:g}) treats the
deviations of the PSF from Gaussianity perturbatively in real space,
without first measuring moments.  In our tests this was the most
successful
method and the only one that comes close to the precision requirements of 
the future surveys. 
The introduction of nonlinear calculations such as higher-order
perturbation theory or direct matrix inversion can be expected to further
improve the accuracy of the method because the full nonlinear dependence
of measured moments on the PSF is taken into account, however one must be
careful to avoid amplifying noise or
creating unstable estimators.

We organize this section as follows: adaptive moment determination is
discussed in Section \ref{sec:adaptive}.  In Section \ref{sec:ele} (and in
Appendix \ref{sec:conv}),
we explore the generalization of the Laguerre expansion of
\citet{2001astro.ph..5178R} and BJ02 to the case of an
elliptical weight.  In Section \ref{sec:psf1}, we explore the limitations
of
BJ02's analytical correction for PSF dilution; the re-Gaussianization
method is introduced in Section \ref{sec:g}.  We conclude by comparing the
methods in simulations
(Section \ref{sec:sim}).

\subsection{\label{sec:adaptive}Adaptive Moments}

 A substantial literature exists
on approaches using circular
Gaussian-weighted moments for $g$ and $I$.
BJ02 introduced ``adaptive'' moments that use an
elliptical weight function whose shape is matched to that of the
object.  The adaptive
moments of an image $I({\bmath x})$ are
determined using the Gaussian with the least-square deviation from the
image, i.e. we minimize:

\beq
E = {1\over 2} \int_{{\bf R}^2} \left| I({\bmath x}) - A\exp\left[
-{1\over
2} ({\bmath x}-{\bmath x}_0)^T {\bf M}^{-1}
({\bmath x}-{\bmath x}_0) \right] \right|^2 \rmd^2{\bmath x}
\label{eq:energy}
\eeq
over the sextuplet of quantities $(A,{\bmath x}_0,{\bf M})$.  Defining:

\beq
w({\bmath x}) = \exp\left[ -{1\over 2} ({\bmath x}-{\bmath x}_0)^T {\bf
M}^{-1}
({\bmath x}-{\bmath x}_0) \right],
\eeq
we note that minimization of $E$ yields the moments satisfying:

\beqa
{\bmath x}_0 && = \frac{ \int_{{\bf R}^2} {\bmath x}w({\bmath
x}) I({\bmath
x}) \rmd^2{\bmath x} }{
\int_{{\bf R}^2} w({\bmath x}) I({\bmath x}) \rmd^2{\bmath x}
}
\nonumber\\
M_{ij} && = 2 \frac{ \int_{{\bf R}^2} ({\bmath x}-{\bmath x}_0)_i
({\bmath x}-{\bmath
x}_0)_j w({\bmath x}) I({\bmath x}) \rmd^2{\bmath x} }{
\int_{{\bf R}^2} w({\bmath x}) I({\bmath x}) \rmd^2{\bmath x}
}
.
\label{eq:adaptive}
\eeqa
Because of the presence of the $w$, the adaptive moments are
weighted.  [It is simpler to work with unweighted moments, but these have
divergent noise (KSB95).]

The ellipticity of an object is then defined in terms of the
adaptive second moments of $I({\bmath x})$:

\beqa
e^{(I)}_+      && = (M_{xx}-M_{yy})/T^{(I)} \nonumber\\
e^{(I)}_\times && = 2M_{xy}/T^{(I)} \nonumber\\
T^{(I)}        && = (M_{xx}+M_{yy});
\label{eq:edef}
\eeqa
the spin-2 tensor ${\bf e}=(e_+,e_\times)$ is called the ellipticity
tensor and the scalar $T^{(I)}$ is called the
trace.

\subsection{\label{sec:ele}Elliptical Laguerre Expansion}

This section explores aspects of the Laguerre expansion introduced by
BJ02.
In particular, we rework BJ02's formalism in a form useful for studying
elliptical objects.
A Laguerre expansion is constructed around a Gaussian with some symmetric
covariance matrix ${\bf M}$; we begin by linearly
transforming the Cartesian $(x,y)$ coordinate system to another $(u,v)$:
\beq
\left( \begin{array}{c} u \\ v \end{array} \right)
=
{\bf M}^{-1/2}
\left( \begin{array}{c} x \\ y \end{array} \right)
\equiv
{1\over\sqrt\zeta}
\left( \begin{array}{cc} M_{yy}+\sqrt D & -M_{xy} \\ -M_{xy} &
M_{xx}+\sqrt D \end{array} \right)
\left( \begin{array}{c} x \\ y \end{array} \right)
,
\label{eq:udef}
\eeq
where $D = M_{xx}M_{yy}-M_{xy}^2$ is the determinant of ${\bf M}$ and
$\zeta = D({M_{xx} + M_{yy} + 2\sqrt D})$.
This provides us with the property that
$u^2+v^2 = {\bmath x}^T{\bf M}^{-1}{\bmath x}$.
The Laguerre basis functions are:

\beq
\psi_{pq}(u,v) = {1\over\sqrt{\pi D\cdot p!q!}} \rme^{-(u^2+v^2)/2}
\Lambda_{pq}(u,v),
\label{eq:psi1}
\eeq
where:

\beq
\Lambda_{pq}(u,v) = (u\pm \rmi v)^{|m|} \sum_{s=0}^{\min(p,q)} (-1)^s
{p!q!\over s!(p-s)!(q-s)!} (u^2+v^2)^{\min(p,q)-s};
\label{eq:psi2}
\eeq
here $m=p-q$ is the angular momentum quantum number and the $+$ sign is
used in $u\pm iv$ if $m>0$ and the $-$ sign is used if $m<0$.  (Both signs
are equivalent if
$m=0$.)  Equations (\ref{eq:psi1}) and (\ref{eq:psi2}) can be seen to be
equivalent to BJ02's equation (6-8) through the use of the explicit
expression for the associated
Laguerre polynomials.  Note that $\Lambda_{pq}$ is a polynomial in two
variables with complex integer coefficients, and
$\Lambda_{pq}$ is the complex conjugate of
$\Lambda_{qp}$.  In Table \ref{tab:lambda} the $\Lambda$ polynomials are
given through fourth order; in the last column we give the result in polar
coordinates.
We note that the $\Lambda$ polynomials are related to
\citet{2001astro.ph..5178R}'s polar Hermite polynomials
according to $\Lambda_{pq}(r\cos\phi,r\sin\phi) =
H_{p,q}(r) \rme^{im\phi}$.

% Table of Laguerre basis functions
\begin{table}
\caption{\label{tab:lambda}The $\Lambda$ polynomials up to fourth order.}
\begin{tabular}{|r|r|c|c|}
\hline
$p$ & $q$ & $\Lambda_{pq}(u,v)$ & $\Lambda_{pq}(r\cos\phi,r\sin\phi)$ \\
\hline
0 & 0 & $1$ & $1$ \\
\hline
1 & 0 & $u+\rmi v$ & $r\rme^{\rmi\phi}$ \\
0 & 1 & $u-\rmi v$ & $r\rme^{-\rmi\phi}$ \\
\hline
2 & 0 & $u^2-v^2 + 2\rmi uv$ & $r^2\rme^{2\rmi\phi}$ \\
1 & 1 & $u^2+v^2-1$ & $r^2-1$ \\
0 & 2 & $u^2-v^2 - 2\rmi uv$ & $r^2\rme^{-\rmi\phi}$ \\
\hline
3 & 0 & $u^3-3uv^2+\rmi(3u^2v-v^3)$ & $r^3\rme^{3\rmi\phi}$ \\  
2 & 1 & $u^3+ uv^2-2u + \rmi(u^2v+v^3-2v)$ & $(r^2-2)r\rme^{\rmi\phi}$ \\
1 & 2 & $u^3+ uv^2-2u + i(-u^2v-v^3+2v)$ & $(r^2-2)r\rme^{-\rmi\phi}$ \\
0 & 3 & $u^3-3uv^2+\rmi(-3u^2v+v^3)$ & $r^3\rme^{-3\rmi\phi}$ \\
\hline
4 & 0 & $u^4 -6u^2v^2 + v^4 + \rmi(4u^3v-4uv^3)$ & $r^4\rme^{4\rmi\phi}$
\\
3 & 1 & $u^4-v^4-3u^2+3v^2+\rmi(2u^3v+2uv^3-6uv)$ &
$(r^2-3)r^2\rme^{2\rmi\phi}$ \\
2 & 2 & $u^4 + 2u^2v^2 + v^4 - 4u^2 - 4v^2 +2$ & $r^4-4r^2+2$ \\
1 & 3 & $u^4-v^4-3u^2+3v^2+\rmi(-2u^3v-2uv^3+6uv)$ &
$(r^2-3)r^2\rme^{-2\rmi\phi}$ \\
0 & 4 & $u^4 -6u^2v^2 + v^4 + \rmi(-4u^3v+4uv^3)$ & $r^4\rme^{-4\rmi\phi}$
\\
\hline
\end{tabular}
\end{table}

Finally, we consider the Fourier transform:

\beq
\tilde f({\bmath k}) = {1\over 2\pi} \int f({\bmath x}) \rme^{-\rmi{\bmath
k}\cdot{\bmath x}} \rmd^2{\bmath x}
\eeq
The Fourier transform of a Laguerre basis function is given by BJ02's
equation (6-62); using our $\Lambda$ polynomials, it is:

\beq
\tilde\psi_{pq}({\bmath k}) = {\rmi^N\over\sqrt{\pi\cdot p!q!}}
e^{-(k_u^2+k_v^2)/2} \Lambda_{pq}(k_u,k_v),
\label{eq:fourier}
\eeq
where $N=p+q$ is the order of the polynomial. 
 The variables $k_u$ and $k_v$ are given by $(k_u,k_v)={\bf
M}^{1/2}{\bmath
x}$ and satisfy $k_uu+k_vv={\bmath k}\cdot{\bmath
x}$.

Now any image can be expanded in the Laguerre basis functions:

\beq
I({\bmath x}) = \sum_{p,q} b_{pq} \psi_{pq}({\bmath x};{\bf M}).
\label{eq:lexpand}
\eeq
(Note that, through an abuse of notation, we are writing $\psi$ now as
though it took ${\bmath x}$ as an argument.)  We
note that there is a complex-conjugate relationship 
among the moments $b_{pq}=b_{qp}^\ast$ if the image $I$ is
real-valued.  Furthermore, by orthogonality of the
$\psi_{pq}$ basis functions, there is an inverse transform 
for the coefficients:

\beq
b_{pq} = \int_{{\bf R}^2} \psi_{pq}^\ast ({\bmath x};{\bf M}) I({\bmath
x}) \rmd^2{\bmath x}.
\eeq
We will find it useful to use the reduced quantities
$\beta_{pq}=b_{pq}/b_{00}$.  BJ02 observe that the Laguerre
coefficients
associated with the adaptively determined covariance ${\bf M}$ satisfy
$\beta_{01}=\beta_{11}=\beta_{02}=0$.  One can
see this directly from the least-squares residual principle by
differentiating equation (\ref{eq:energy}) with respect
to $(A,{\bmath x}_0,{\bf M})$ and setting the result to zero.

\subsection{\label{sec:psf1}PSF Corrections}

PSF correction is typically performed by correcting the moments of the
measured object using the moments of the
PSF.  
  There are
generally two cases to consider here: either 
the PSF moments are
measured with an elliptical Gaussian weight adapted to the PSF shape
rather than the object shape, or they
are measured with the same weight function as the object.  The former
approach is used by BJ02, and we will consider it
in detail here; we defer investigation of PSF moments determined with the
same weight function as the object.

Another issue that arises is whether to use a rounding-kernel method to
symmetrize the PSF.  Here we assume that the best-fit Gaussian to
the PSF is circular; this could be achieved using a rounding
kernel.  An alternative is to apply a shear (to both the image and PSF)
to make the best-fit Gaussian to the PSF circular, and then apply an
inverse shear after making the ellipticity measurement.  [The
transformation of ellipticities is performed according to equation
(\ref{eq:composition}).]  The rounding kernel method may be better
 suited to
elimination of spurious power from PSF anisotropy, since a rounding
kernel can be designed to eliminate arbitrarily many spin-2 moments
of the PSF (at least in principle, see e.g. BJ02); however, rounding
kernels increase the effective PSF size which is not desirable.
Circularizing the best-fit Gaussian to the PSF with a shear does not
increase PSF size, but the removal of spin-2 moments of the PSF
(and hence of the associated spurious power) is incomplete as 
only the $b_{02}$
moment is forced to zero by this method.

If the galaxy and PSF are both elliptical Gaussians, the PSF correction
is trivial because the covariance matrix of the galaxy and PSF add to
yield the covariance of the
observed image.  In this case, we have ${\bf M}_I={\bf M}_f + {\bf M}_g$
and hence:

\beqa
e^{(f)}_+ = \frac{
({\bf M}_I-{\bf M}_g)_{xx} - ({\bf M}_I-{\bf M}_g)_{yy}
}{
({\bf M}_I-{\bf M}_g)_{xx} + ({\bf M}_I-{\bf M}_g)_{yy}
}
\nonumber\\
e^{(f)}_\times = \frac{
2 ({\bf M}_I-{\bf M}_g)_{xy}
}{
({\bf M}_I-{\bf M}_g)_{xx} + ({\bf M}_I-{\bf M}_g)_{yy}
}
.
\label{eq:gausscorr}
\eeqa
If we were using unweighted moments, this scheme would be valid for any
PSF and galaxy.
For weighted moments, equation (\ref{eq:gausscorr}) fails, and a
correction for non-Gaussianity of the galaxy and PSF is necessary.

BJ02's prescription (see their Appendix C) for PSF corrections is 
as follows: begin by applying
a rounding kernel; then the ellipticity of the
measured object
${\bf e}^{(I)}$ is transformed into the ellipticity of the intrinsic
object according to ${\bf e}^{(f)}={\bf
e}^{(I)}/R$, where:

\beq
R = 1 - { T_g(1-\beta^{(g)}_{22})/(1+\beta^{(g)}_{22}) \over
T_I(1-\beta^{(I)}_{22})/(1+\beta^{(I)}_{22}) } .
\label{eq:bjr}
\eeq
This equation is designed to reduce to the correct form in the limit of
well-resolved objects with Gaussian PSFs,
and in the limit of any Gaussian PSF and Gaussian object (whether or not
it is well-resolved,
see Appendix C of BJ02); however, no rigorous derivation has been
provided.  Indeed, we can
show that for non-Gaussian PSFs, the above equation may be subject to
significant error.
The measured object has intensity:

\beq
I({\bmath x}) = \int_{{\bf R}^2} f({\bmath x}') g({\bmath x}-{\bmath
x}') \rmd^2{\bmath x}'
=
\sum_{n=0}^\infty {1\over n!} q_{i_1...i_n} {\partial^n\over\partial
x_{i_1}...\partial x_{i_n}} f({\bmath x})
,
\label{eq:psfseries}
\eeq
where

\beq
q_{i_1...i_n} = \int_{{\bf R}^2} g({\bmath x}) x_{i_1}...x_{i_n}
\rmd^2{\bmath x}
\label{eq:q}
\eeq
is the unweighted $n$th moment of the PSF.  (Note that formally $q$ is
infinite.)  In the limit of a well-resolved object, the effect of the PSF
is dominated by the second
($n=2$) moment because the series in equation
 (\ref{eq:psfseries}) converges rapidly
and the first moment vanishes (we assume without loss of generality that
the centroid of the PSF
lies at the origin).  In this limit, BJ02 showed that:

\beq
R = 1 - { q_{xx}+q_{yy} \over
T_I(1-\beta^{(I)}_{22})/(1+\beta^{(I)}_{22}) } + O(T_g^2/T_I^2).
\eeq
The relation between $q_{xx}+q_{yy}$ and the Laguerre moments can be
computed:

\beq
q_{ij} = \int_{{\bf R}^2} g({\bmath x}) x_i x_j \rmd^2{\bmath x}
=
\frac{
\sum_{pq} (p!q!)^{-1/2} \beta_{pq} \int_{{\bf R}^2} x_i x_j
\Lambda_{pq}({\bf M}^{-1/2}{\bmath x}) \rme^{-{\bmath x}{\bf
M}^{-1}{\bmath x} /
2} \rmd^2{\bmath x}
}{
\sum_{pq} (p!q!)^{-1/2} \beta_{pq} \int_{{\bf R}^2} \Lambda_{pq}({\bf
M}^{-1/2}{\bmath x}) \rme^{-{\bmath x}{\bf M}^{-1}{\bmath x} / 2}
\rmd^2{\bmath x}
}
= M_{ij} [ 1 + 4\beta_{22}^{(g)} ] + ... 
,
\eeq
where ``...'' represents both higher-order terms in $\beta_{22}^{(g)}$
and terms from the higher-order Laguerre moments $b_{pq}^{(g)}$ with
$p>2$ or $q>2$.  One would thus
expect that for well-resolved galaxies,
the
correction for non-Gaussianity of the PSF should take the form:

\beq
R = 1 - {T_g\over T_I} \left[
\frac{ 1 + 4\beta_{22}^{(g)} + ... }{
(1-\beta^{(I)}_{22})/(1+\beta^{(I)}_{22}) }  \right]
+ O(T_g^2/T_I^2),
\eeq
where the numerator of the quantity in brackets is valid to
 first order in $\beta_{22}^{(g)}$ and to zeroeth order in
the higher moments.
 We note that the BJ02 correction does not exhibit this behavior; its
numerator has
the form $1 - 2\beta_{22}^{(g)}$ instead of $1 +
4\beta_{22}^{(g)}$.  Thus we expect that for PSFs with positive
kurtosis  $\beta_{22}^{(g)}>0$, equation
(\ref{eq:bjr}) will overestimate the resolution factor and hence
underestimate the shear.  This expectation is verified in Section
\ref{sec:sim}.  Note additionally that the above argument does not
apply for the less well-resolved galaxies (i.e. those for which
$T_g$ is comparable to $T_I$); in this case, the factor of
$1 + 4\beta_{22}^{(g)}$ is not correct.  The correct factor is
worked out in Appendix \ref{sec:psfcorr}.

Several methods could be used to improve equation (\ref{eq:bjr}).
One approach is to compute the correction to the galaxy ellipticity to
linear order in the
$\beta_{pq}$; this approach is pursued in Appendix
\ref{sec:psfcorr}.  Unfortunately, realistic PSFs (e.g. turbulence-
or diffraction-limited PSFs) have significant power
at large radii and hence the convergence of the Laguerre expansion is
rather slow.  This is manifested in Section \ref{sec:sim}, where
significant errors are found from
terminating the Laguerre expansion after the 22 moment.  
Therefore, in the next section we will pursue an alternative approach: we
fit a Gaussian to
the PSF, compute the residuals,
and attempt to correct the galaxy image for the effects of these
residuals.  Then we apply equation (\ref{eq:bjr}) to this corrected
galaxy image.  As shown in the
simulations (Section \ref{sec:sim}), this method yields good performance
for non-Gaussian PSFs and galaxies, except for the most elongated
galaxies.

\subsection{\label{sec:g}Re-Gaussianization}

Let us suppose that the PSF $g$ can be approximated by a Gaussian $G$
with some covariance ${\bf M}_G$:

\beq
g({\bmath x}) \approx G({\bmath x}) = {1\over 2\pi\sqrt{\det{\bf M}_G}}
\exp
\left( -{1\over 2} {\bmath x}^T{\bf M}_G^{-1}{\bmath x} \right).
\label{eq:gdef}
\eeq
We may then define the residual function $\epsilon({\bmath x}) = g({\bmath
x}) - G({\bmath x})$.  It follows that the measured image
intensity will satisfy
$I = g \otimes f = G \otimes f + \epsilon \otimes f$,
where $\otimes$ represents convolution.  This can be rearranged to yield:

\beq
G \otimes f = I - \epsilon \otimes f.
\label{eq:gf}
\eeq
This equation thus allows us to determine the Gaussian-convolved
intrinsic galaxy image $I'$, if we know $f$.  At first glance this
does not appear helpful, since if we knew $f$ it would be trivial to
determine $G\otimes f$.  However, $f$ appears in this
equation multiplied by (technically, convolved with) a small correction
$\epsilon$, so equation (\ref{eq:gf}) may be reasonably
accurate even if we use an approximate form for $f$.  The simplest
approach is to approximate $f$ as a Gaussian with covariance:

\beqa
f_0 && = {1\over 2\pi\sqrt{\det{\bf M}_f^{(0)}}} \exp \left( -{1\over 2}
{\bmath x}^T[{\bf M}_f^{(0)}]^{-1}{\bmath x} \right),
\nonumber \\
{\bf M}_f^{(0)} && = {\bf M}_I - {\bf M}_g,
\label{eq:mf0}
\eeqa
where ${\bf M}_I$ and ${\bf M}_g$ are the adaptive covariances of the
measured object and PSF, respectively.  Then we can define:

\beq
I' \equiv I - \epsilon \otimes f_0 (\approx G\otimes f).
\label{eq:iprime}
\eeq
The adaptive moments of $I'$ can then be computed, and the PSF correction
of equation (\ref{eq:bjr}) can then be applied to recover
the intrinsic ellipticity ${\bf e}^{(f)}$.

There are several ``knobs'' that can be adjusted on the
re-Gaussianization method.  First, there is the choice of Gaussian
approximation to the PSF, equation (\ref{eq:gdef}).  We have chosen to
set ${\bf M}_G$ to be the adaptive covariance of the PSF,
i.e. the covariance of the best-fit Gaussian.  However, rather than using
the normalization of the best-fit Gaussian, we have
normalized equation (\ref{eq:gdef}) to integrate to unity.  While this
increases the overall power $\int(\epsilon^2)$ of the
residual function, it yields $\int\epsilon=0$, which ensures that for
well-resolved objects (i.e. objects for which the PSF is
essentially a $\delta$-function), the ``correction'' $\epsilon\otimes
f_0$ applied by equation (\ref{eq:iprime}) does not corrupt
the image $I$.  One could also try a more sophisticated function $f_0$
than a Gaussian, perhaps based on a partial deconvolution of
the PSF; we have not explored the possibilities here, but it is clear
that an attempt to approximate $f$ will have to make some
assumptions regarding the behavior of $f$ in the regions of Fourier space
where the PSF has little power (i.e. where the optical
transfer function is close to zero).

\subsection{\label{sec:sim}Simulation}

\begin{figure}
\includegraphics[angle=-90, width=6.8in]{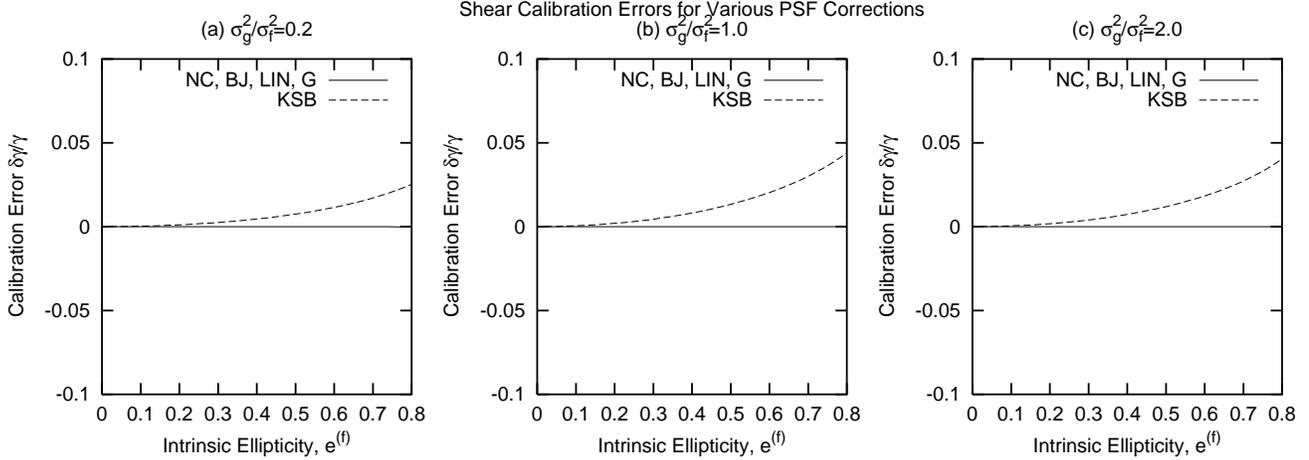}
\caption{\label{fig:sim0}The error in reconstructing the shear for the
case of a circular Gaussian PSF and elliptical Gaussian galaxy.
  We have
plotted the shear calibration error $\delta\gamma /\gamma$.  Note that
the adaptive moments methods are exact for this case.  The KSB95 method is
not exact but gives
good
 performance (calibration errors of order 1 per cent for most
galaxies) even when the galaxy is not well resolved.
}
\end{figure}

\begin{figure}
\includegraphics[angle=-90, width=6.8in]{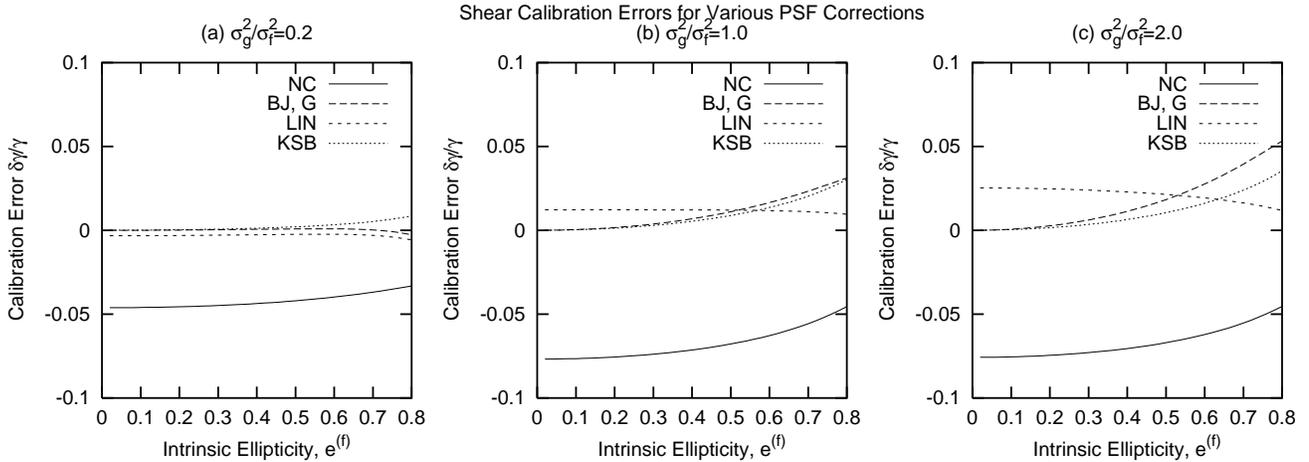}
\caption{\label{fig:sim1}The error in reconstructing the shear for the
case of a circular Gaussian PSF and elliptical exponential galaxy.  We
have
plotted the shear calibration error $\delta\gamma /\gamma$.  Note that
for the well-resolved galaxy (a), both BJ02 and linear methods provide
good performance; the error
in the linear method grows substantially for the poorly resolved galaxies
in (b) and (c).  The no-correction method underestimates ellipticities in
all cases, with the
situation worsening for the poorly resolved galaxies.
}
\end{figure}

\begin{figure}
\includegraphics[angle=-90, width=6.8in]{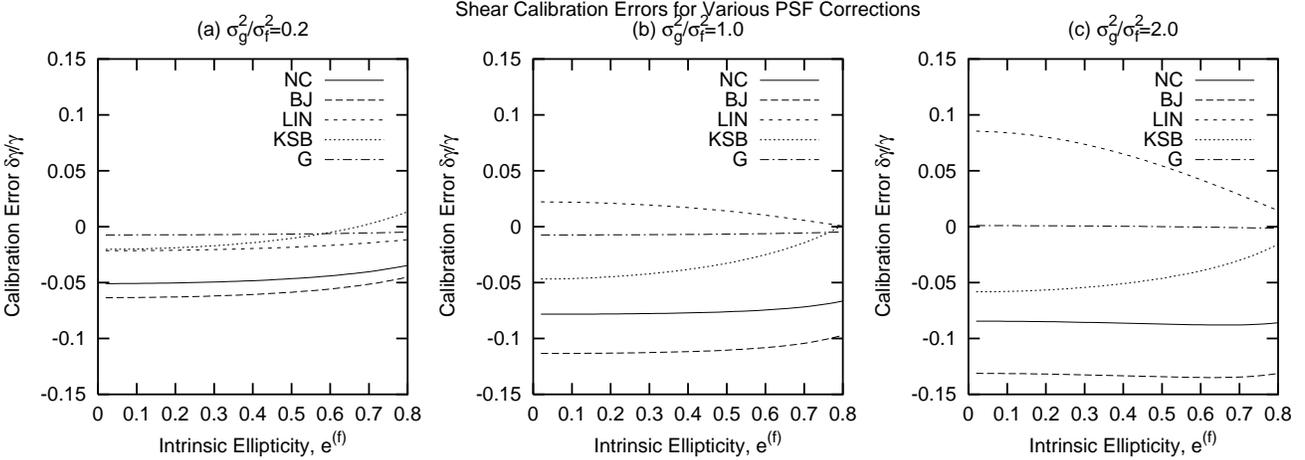}
\caption{\label{fig:sim2}The error in reconstructing the shear for the
case of a circular PSF generated by atmospheric turbulence and an
elliptical
Gaussian galaxy.  We have 
plotted the shear calibration error $\delta\gamma /\gamma$.
For well-resolved (a) and moderately-resolved (b) galaxies, the linear
method substantially reduces errors when compared to the no-correction
method.  For poorly resolved
galaxies, the linear method over-corrects for PSF non-Gaussianity.  The
PSF non-Gaussianity correction of BJ02 has the incorrect sign in all
three cases.
Note that the scale is different from that of Fig. \ref{fig:sim1}.
}
\end{figure}

We may test the PSF correction algorithms via simulation for common
categories of galaxy and PSF.  The simulation
procedure is as follows:

\begin{list}{$\bullet$}{}
\item Construct an intrinsic image $f$ of the test galaxy;
\item Construct a sample PSF $g$;
\item Convolve the test galaxy and PSF, then pixelize this to yield a
``measured object'' $I$;
\item Compute the adaptive moments of the test galaxy $f$, and measured
galaxy $I$;
\item Apply the PSF correction scheme to the moments of $I$ to yield the
``recovered'' ellipticity of the test galaxy;
\item Compare the true and recovered ellipticities ${\bf e}^{(f)}$ of the
test galaxy;
\item Propagate the ellipticity error through to an error in the shear
estimate.
\end{list}

Our simulations here will compare five methods of recovering the initial
ellipticity.  The re-Gaussianization (``G'') method is that described in
Section \ref{sec:g}.  
The linear (``LIN'') method follows our prescription at the end of
Appendix
\ref{sec:psfcorr}.
The ``BJ'' method uses BJ02's resolution factor from Appendix C,
 equation (\ref{eq:bjr}),
instead of ours, equation (\ref{eq:4corr2}).  [BJ02 discuss several
methods of correcting for PSF dilution, including a matrix inversion
method which from theoretical expectations should do better than the
resolution factor approach.  The resolution factor used here is the
same one that is used in the cosmic shear study of
\citet{2002astro.ph.10604J} and can be implemented on Sloan Digital
Sky Survey PHOTO reductions.]
  The no-correction
(``NC'') method
computes ellipticities based on equation (\ref{eq:gausscorr}) and hence
is correct only for Gaussian galaxies and PSFs.
Included for comparison is the ``KSB'' method, which uses the KSB95
method with pre-seeing shear polarizability
estimated by the method of \citet{1997ApJ...475...20L}.
Note that there are several versions of the KSB method in the literature, 
a comparison of some of these is provided by \citet{2001A&A...366..717E}.
  (For details of
our
implementation of KSB95, see Appendix \ref{sec:ksb}.)

As our objective is to correct for the dilution
and smearing of the galaxy image by the PSF, we do not include noise in
these simulations; thus noise-rectification
biases such as \citet{2000ApJ...537..555K}'s ``noise bias'' (termed
``centroid
bias'' in BJ02) are not addressed here and a separate correction for
these must be made.  The bias in the PSF-corrected ellipticity ${\bf e}$ 
introduced by noise is generally given by the Taylor expansion:

\begin{equation}
\langle\Delta{\bf e}\rangle = \frac{1}{2}\frac{\partial^2{\bf e}}{\partial 
I_i \partial I_j} \langle \Delta I_i \Delta I_j \rangle + ...,
\label{eq:noise}
\end{equation}
where $I_i$ is the flux in pixel $i$.  Here the leading order term scales 
with the significance $\nu$ as $1/\nu^2$, and the higher-order terms scale 
as $1/\nu^3$, $1/\nu^4$, etc.  For galaxies detected at high significance, 
the dominant noise bias contribution is from the leading $1/\nu^2$ term 
(shape measurement of low-significance galaxies is unstable and not 
recommended).  Present approaches to the evaluation of the noise bias 
include analytic solutions for Gaussian objects 
\citep{2000ApJ...537..555K} 
and analytic scaling relations whose coefficients can be determined 
through simulations (BJ02).  There is also a Monte Carlo approach, valid 
for any object profile and any PSF method: if we add artificial Gaussian 
noise to the measured image with covariance $\zeta\langle\Delta I_i\Delta 
I_j\rangle$ (where the parameter $\zeta\ll 1$), and then take an 
ensemble-averaged change in measured ellipticity over many 
realizations of the artificial noise, we find the 
ensemble-averaged change in measured ellipticity $\hat{\bf 
E}=\zeta\langle\Delta{\bf e}\rangle$.  This allows us to determine 
$\langle\Delta{\bf e}\rangle$, and thus correct the measured ellipticity 
for noise bias: ${\bf e}\mapsto{\bf e}-\zeta^{-1}\hat{\bf E}$.  Note that 
the statistical variance in ${\bf e}$ resulting from the Monte Carlo 
method after $N$ realizations of the artificial noise have been computed 
is $\langle\Delta{\bf e}^2\rangle/N$ whereas the statistical noise in the 
galaxy shape measurement is $\langle\Delta{\bf e}^2\rangle$; hence the 
Monte 
Carlo convergence rapidly becomes a negligible contributor to the overall 
statistical uncertainty.  Thus we can determine and remove the noise bias 
effect on any PSF correction method by evaluating Eq. (\ref{eq:noise}), 
provided only that the significance $\nu$ is large enough to treat the 
noise as a perturbation.  Since the $O(1/\nu^2)$ noise bias is 
straightforward to remove and in any case the details of its removal 
process do not depend strongly on the PSF correction scheme, we have 
presented the simulations here without noise to illustrate the purely 
PSF-induced effects.

Since the end result of the shape measurements is to produce a shear
estimate, we show the calibration error $\delta\gamma/\gamma$ in the
shear estimator as a function of
object and PSF parameters.  For the adaptive moment methods (no
correction, linear, BJ02) that yield ellipticity estimates for individual
objects, we proceed as
follows: consider the (unweighted) shear estimator:

\beq
\hat\gamma_+ = \frac{\langle \hat e_+ \rangle}{2 \left< 1-{1\over 2}\hat
e^2 \right> }
,
\eeq
where $\hat {\bf e}$ is the measured ellipticity.  For objects of a fixed
ellipticity this has calibration:

\beq
\left.
{\partial\hat\gamma_+\over\partial\gamma_+} \right|_{\gamma=0} = {1\over
2-\hat e^2}\left< \frac{ \partial \hat e_+}{ \partial \gamma_+ } \right>.
\eeq
Now if we use the transformation law for ellipticities under
infinitesimal shear, $e_\alpha\mapsto e_\alpha + 2\gamma_\alpha -
2e_\alpha e_\beta \gamma_\beta$ [see
our equation (41) in the $\delta\rightarrow 0$ limit, and note that the
infinitesimal shear $\gamma$ differs from that of the distortion shear
$\delta$ by a factor of
two], we can azimuthally average to yield:

\beq
\left.
{\partial\hat\gamma_+\over\partial\gamma_+} \right|_{\gamma=0}
= {1\over 2-\hat e^2}\left( (1-e^2){\partial \hat e\over\partial e} +
{\hat e\over e} \right)
.
\label{eq:calib_e}
\eeq
This is the equation we use to determine the calibration for the no
correction, linear, and BJ02 methods.  In the plots
(Figs. \ref{fig:sim0}--\ref{fig:sim2}) we show
the error $\delta\gamma/\gamma =
{\partial\hat\gamma_+\over\partial\gamma_+} |_{\gamma=0} - 1$.

% Table of PSF forms
\begin{table}
\caption{\label{tab:psf}PSFs used in the simulations.  The adaptive
(i.e. iterated until $\beta_{11}=0$) radial $2n$th moments $\beta_{nn}$
characterize the
non-Gaussianity of the PSF.  Note the slow convergence of the Laguerre
expansion for the diffraction-limited PSF.}
\begin{tabular}{clclclrlrlrlr}
\hline
PSF type & & equation & & adaptive width, $\sigma$ & & $\beta_{22}$~ & &
$\beta_{33}$~ & & $\beta_{44}$~ & &
$\beta_{55}$~\\
\hline
Gaussian & & eq. (\ref{eq:psfgauss}) & & $\sigma$ & & $0.000$ & & $0.000$
& & $0.000$ & & $0.000$ \\
turbulence-limited & & eq. (\ref{eq:psfturbulent}) & & $0.45\theta_\FWHM$
& & $+0.046$ & & $+0.012$ & & $+0.010$ & &
$+0.006$ \\
Airy (diffraction-limited) & & eq. (\ref{eq:psfdiffraction}) & &
$0.41\lambda/D$ & & $-0.043$ & & $+0.023$ & & $+0.039$
& & $+0.022$
\\
quartic-Gaussian & & eq. (\ref{eq:qg}) & & $\sigma$ & & $+0.046$ & &
$0.000$ & & $0.000$ & & $0.000$ \\
\hline
\end{tabular}
\end{table}

For the KSB95 method, equation (\ref{eq:calib_e}) is not applicable
because
the method does not produce an estimator for the adaptive
ellipticity.  Rather it produces a
non-adaptive ellipticity ${\bf e}^K$ and a pre-seeing shear
polarizability tensor $P^\gamma$.  The shear is then estimated via:

\beq
\hat\gamma_\alpha = [\langle P^\gamma \rangle ^{-1}]_{\alpha\beta}
\langle e^K_\beta \rangle.
\eeq
For an ensemble of galaxies of the same ellipticity but rotated by
different angles, $\langle P^\gamma \rangle$ becomes isotropized and is
equal to ${1\over 2}\Tr
P^\gamma$.  We can find $\langle {\bf e}^K\rangle $ by an argument
similar to that leading to equation (\ref{eq:calib_e}); the result is:

\beq
\left. {\partial\hat\gamma_+\over\partial\gamma_+} \right|_{\gamma=0} =
 {2\over\Tr P^\gamma} \left( (1-e^2){\partial e^K\over\partial e} +
{e^K\over e} \right)
.
\label{eq:calib_k}
\eeq
This is the calibration equation that we use for the KSB95 method.

We use several different forms for the PSF in our simulations; these are
given in Table \ref{tab:psf}.  The first is the Gaussian
form, with PSF $g(r)$ and optical transfer function (OTF) $\bar g(k)$
given by:

\beqa
g(r) && = {1\over 2\pi\sigma^2} \rme^{-r^2/2\sigma^2},\nonumber\\
\bar g(k) \equiv 2\pi\tilde g(k) = \int_{{\bf R}^2} \rme^{-\rmi{\bmath
k}\cdot{\bmath x}} g({\bmath x}) \rmd^2{\bmath x} && =
\rme^{-\sigma^2k^2/2}.
\label{eq:psfgauss}
\eeqa
(We define the OTF $\bar g(k)$ to be $2\pi$ times the Fourier transform
$\tilde g(k)$ of the PSF so that $\bar g(0)=1$.)
The Gaussian, while simple to analyze, is not a good approximation to the
PSF for real experiments.  For ground-based experiments,
the resolution is generally limited by atmospheric turbulence in the
inertial range; the OTF for this case is given by
(see \citealt{2000ApJ...537..555K},
Appendix A):

\beqa
g(r) && = \frac{3k_0^2}{10\pi} \sum_{j=0}^\infty (-1)^j
\frac{ \Gamma({6\over 5}(j+1)) }{ j!^2}
 \left( \frac{k_0r}{2} \right)^{2j},
 \nonumber\\
\bar g(k) && = \exp \left[ -(k/k_0)^{5/3} \right],
\label{eq:psfturbulent}
\eeqa
where $k_0=2.92/\theta_\FWHM$ and the seeing $\theta_\FWHM$ is the
full-width half-maximum width of the PSF; typically
this is of
order 1 arcsec.  [The series for $g(r)$ is convergent for all radii,
nevertheless the summation is numerically
unstable at $k_0r\gg 1$ and for computational purposes it is better to
take the Fourier transform of $\bar g(k)$.]
For space-based experiments, the telescope is usually diffraction-limited
and hence the OTF is the autocorrelation
function of the aperture. If the aperture is
circular with diameter $D$, this is:

\beqa
g(r) && = {1\over\pi} \left[ \frac{ J_1(k_0r/2)}{r} \right]^2, \nonumber\\
\bar g(k) && = \left\{ \begin{array}{ll} {2\over\pi} \left[ \arccos
(k/k_0) - (k/k_0)\sqrt{1-k^2/k_0^2} \right] &
,\; k<k_0 \\
0 & ,\; k\ge k_0 \end{array} \right.
,
\label{eq:psfdiffraction}
\eeqa
where $k_0 = 2\pi D/\lambda$
and  $\lambda$ is the wavelength of observation.  We do not pursue an
analysis of the effects of chromaticity, but note
that
(particularly for the diffraction-limited PSF, which has a steeper
wavelength dependence than the turbulence-limited
PSF) the effect may be important.

\begin{figure}
\includegraphics[angle=-90, width=6.8in]{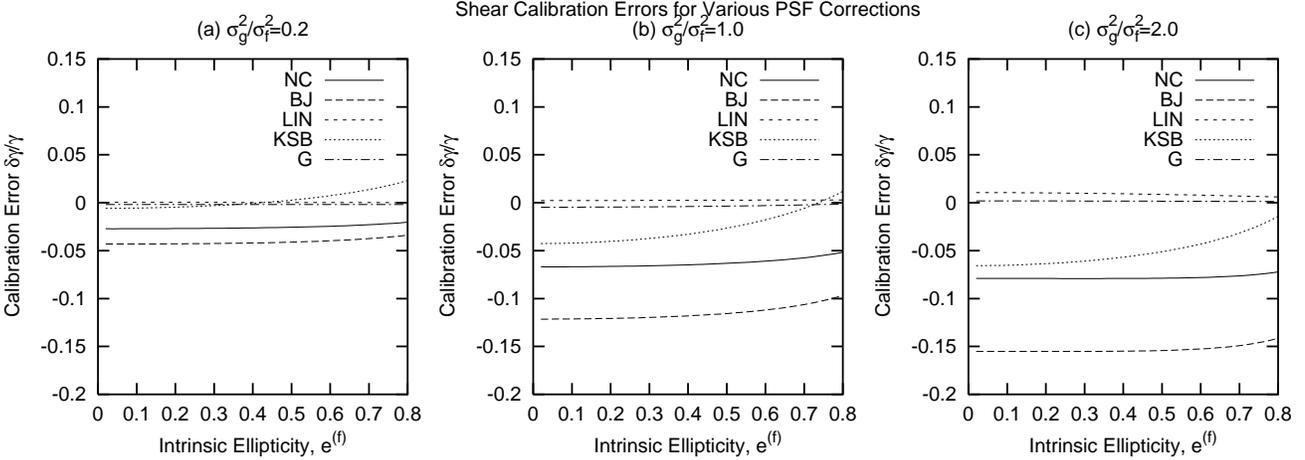}
\caption{\label{fig:sim3}The error in reconstructing the shear for the
case of a circular quartic-Gaussian PSF [see equation (\ref{eq:qg})] and
an elliptical
Gaussian galaxy.  We have 
plotted the shear calibration error $\delta\gamma /\gamma$.
For well-resolved (a) and moderately-resolved (b) galaxies, the linear
method substantially reduces errors when compared to the no-correction
method.  For poorly resolved galaxies, the linear method over-corrects
for PSF non-Gaussianity.  The PSF non-Gaussianity correction of BJ02 has
the incorrect sign in all three cases.
}
\end{figure}

\begin{figure}
\includegraphics[angle=-90, width=6.8in]{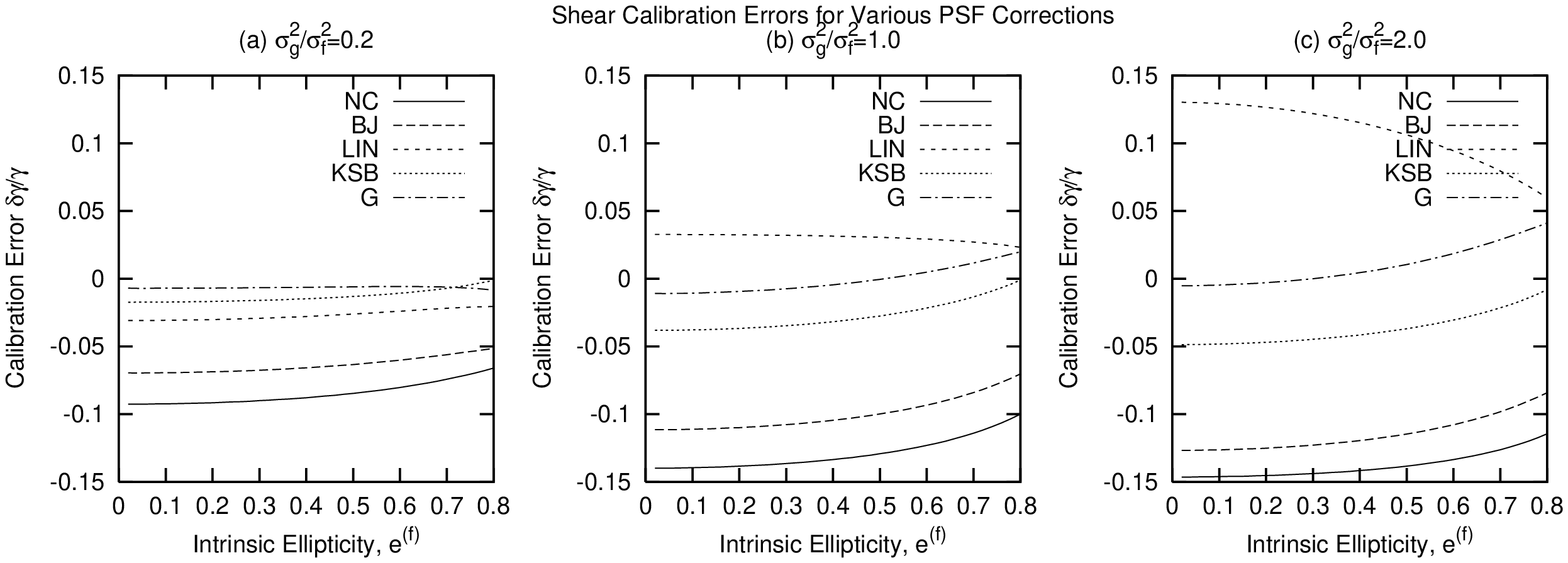}
\caption{\label{fig:sim4}The error in reconstructing the shear for the
case of a circular turbulence-limited PSF and an elliptical
exponential galaxy.  We have 
plotted the shear calibration error $\delta\gamma /\gamma$.  Note the
significant loss of accuracy of all the methods except for
re-Gaussianization (G) for poorly-resolved
galaxies.
}
\end{figure}

\begin{figure}
\includegraphics[angle=-90, width=6.8in]{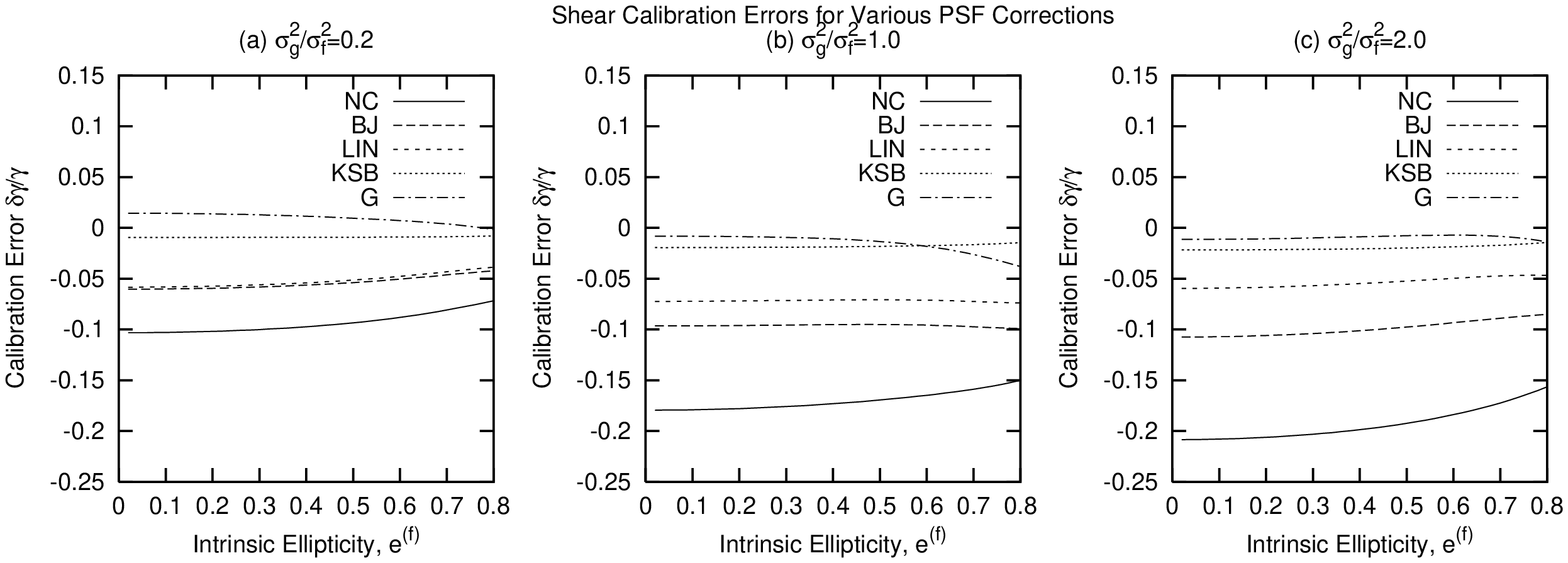}
\caption{\label{fig:sim6}The error in reconstructing the shear for the
case of a circular turbulence-limited PSF and an elliptical 
de Vaucouleurs profile galaxy.  We have 
plotted the shear calibration error $\delta\gamma /\gamma$.  Note the
significant loss of accuracy of all the methods except for
re-Gaussianization (G) and KSB for poorly-resolved galaxies.
}
\end{figure}

\begin{figure}
\includegraphics[angle=-90, width=6.8in]{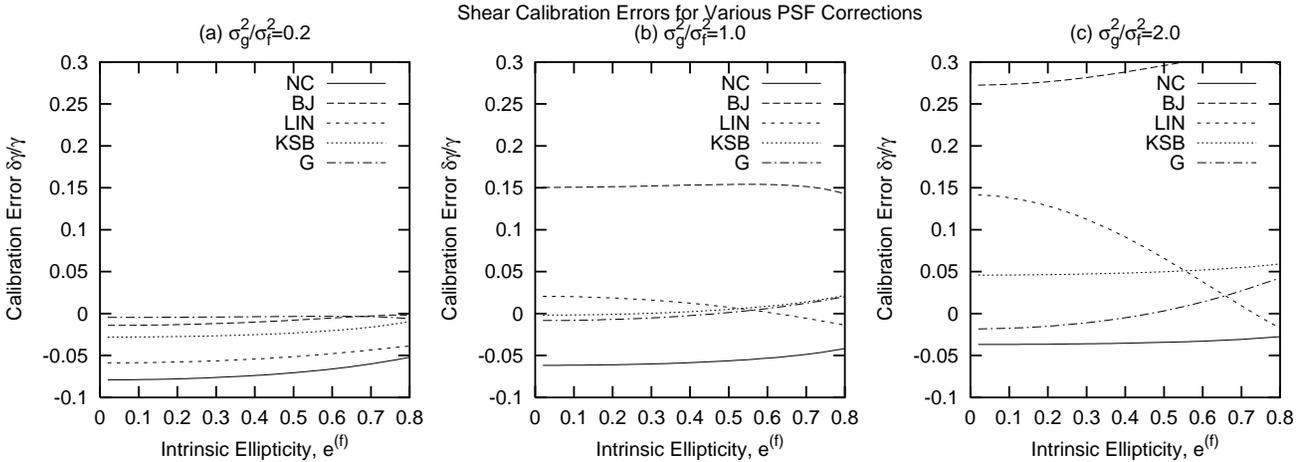}
\caption{\label{fig:sim5}The error in reconstructing the shear for the
case of a circular diffraction-limited PSF and an elliptical
exponential galaxy.  Severe loss of accuracy occurs for most of the
methods if the galaxy is poorly resolved
(c); fortunately, diffraction-limited PSFs occur only for 
space-based surveys in which the PSF size is usually small compared to
the size of the galaxy, hence the comparatively
large errors exhibited in (c) are unlikely to be a problem in practice.
}
\end{figure}

Simulation results are shown in the figures for three cases: a
well-resolved galaxy ($\sigma^2_g/\sigma^2_f = 0.2$; we remind the reader
that subscript 
$g$ is of PSF and $f$ for galaxy), an intermediate case
($\sigma^2_g/\sigma^2_f=1.0$),
and a poorly resolved galaxy
($\sigma^2_g/\sigma^2_f=2.0$).  [Here $\sigma^2\equiv\sqrt{\det\bf M}$ is
proportional to the area covered by the object.]
The simple case of a Gaussian galaxy and Gaussian PSF is shown in
Fig. \ref{fig:sim0}.  The adaptive moments methods are exact for this
case.  The KSB95 method, while
not exact, has relatively small calibration error except for the very
elongated galaxies.

In
Fig. \ref{fig:sim1}, we test the ability of these methods to handle
non-Gaussian galaxies by showing simulation results with a circular
Gaussian PSF and an elliptical
galaxy with
exponential ($\propto \rme^{-r/r_0}$) radial profile.  The exponential
profile has a positive kurtosis $\beta^{(f)}_{22} \approx 0.17$.  Note
that both the BJ02 and linear
methods improve upon the no-correction method; BJ02's performance here is
superior to that of the linear method, presumably because BJ02 works to
all orders in
$\beta^{(f)}_{22}$.  Once again, the KSB95 method is accurate at the 1 per
cent level except for very eccentric galaxies.

In Fig. \ref{fig:sim2}, we display a simulation using an elliptical
Gaussian galaxy and a non-Gaussian PSF.  The PSF is the Fourier transform
of an $\rme^{-Ck^{5/3}}$
function, as
appropriate for observations through a turbulent atmosphere with
Kolmogorov scaling
\citep{2000ApJ...537..555K}.  This PSF has a kurtosis of
$\beta^{(g)}_{22} \approx 0.046$; it is
important to note,
however, that the PSF drops to zero slowly at large radii and hence it
 possesses higher moments, e.g. $\beta^{(g)}_{33}\approx 0.012$ (see Table
\ref{tab:psf}) that are
not taken into account by
either the BJ02 or linear methods.  We can demonstrate that the error
from the linear method is due to these higher moments by repeating the
simulation using a PSF of the
``quartic-Gaussian'' form:

\beq
g(r) = \left( 1 + 0.046\cdot \frac{ r^4/\sigma^4-4r^2/\sigma^2+2}{2}
\right) \rme^{-r^2/2\sigma^2},
\label{eq:qg}
\eeq
which has the same kurtosis as the turbulent PSF but no higher-order
moments.  The results using this form for the PSF are shown in
Fig. \ref{fig:sim3}.  Finally, we
have shown simulations where both the galaxy and PSF are
non-Gaussian: one with a turbulent PSF and exponential galaxy
(Fig. \ref{fig:sim4}), one with 
a turbulent PSF and de Vaucouleurs profile (Fig. \ref{fig:sim6}) and one
with a
diffraction-limited (Airy) PSF and exponential galaxy
(Fig. \ref{fig:sim5}).  Of the methods described, only the
re-Gaussianization method retains
high accuracy in these cases.  This is primarily because of the
sixth-order and higher moments of these PSFs, which are ignored by the
BJ02 and fourth-order linear
methods.  For well-resolved objects, the KSB95 method avoids these
problems
because its non-adaptive weight function allows it to ``see'' the
turbulence-induced tails and/or
diffraction rings in the PSF.  For the less well-resolved objects, KSB
suffers from loss of higher-moment information, although it is apparent
from the simulations that
the problem is not as serious as for the BJ02 or linear methods.

The simulations indicate significant (at least several
percent) calibration errors for all of the currently used
methods when the PSF is non-Gaussian and has size comparable to that of
the galaxy.  Both the KSB95 and BJ02
methods yield an underestimate of the shear for turbulence-limited PSFs,
with the error for poorly resolved objects
being of the order of 5 per cent for KSB95 and 10--15 per cent for BJ02.
The errors have the opposite sign for a diffraction-limited PSF, however
the small size (typically of order 0.1
arcsec) of diffraction-limited PSFs helps to reduce the error in this
case (compare Fig. \ref{fig:sim5}a-c).
The re-Gaussianization method introduced here shows promise in reducing 
these calibration errors to the 1--2 per cent level even for poorly
resolved 
galaxies. It has the best performance of all the methods studied here. 
While we have not simulated \citet{2001astro.ph..5179R}'s matrix
inversion method, we note for comparison that in
realistic Monte Carlo
simulations
(i.e. including noise and pixelization), \citet{2001astro.ph..5179R}
found calibration errors of $-3\pm4$ per cent in $e_+$ and
$0\pm4$ per cent in
$e_\times$ using a turbulent PSF with $\sigma_g^2/\sigma_f^2=0.32$. This 
would thus be comparable to our tests of well resolved galaxies, for 
which KSB95 and re-Gaussianization methods are accurate to within 2 per
cent.
It would be interesting to investigate the behavior
of the matrix inversion method at lower resolution where 
the other methods
except for re-Gaussianization show more significant error.

\section{Shear Selection Bias}
\label{sec:ssbsec}

Even if the PSF correction is perfect, an estimate of the shear can be
biased if the object detection algorithm
preferentially selects galaxies elongated in one direction over galaxies
elongated in another
direction.  \citet{2000ApJ...537..555K} and BJ02
investigate such a selection bias that results from asymmetries of the
PSF and offer methods for correcting
it.  However, there is another selection bias that exists: suppose that
we are looking at a part of the sky where the
gravitational shear elongates objects in the $x$-direction
(i.e. $\gamma_+>0$).  Then background galaxies
intrinsically (i.e. pre-shear) elongated in
the $x$-direction appear more elliptical than otherwise identical
background galaxies elongated in the $y$-direction
(although both have the
same area, flux, and surface brightness).  Many object detection
algorithms, such as significance cutoffs, will
preferentially detect the circular objects in this situation, thus the
catalog of background galaxies that results will
preferentially contain galaxies aligned orthogonal to the gravitational
shear.  This results in underestimation of the
lensing signal.  We will call this effect ``shear selection bias'' to
distinguish it from the ``PSF selection bias''
discussed by \citet{2000ApJ...537..555K} and BJ02.

Unlike PSF-related biases, shear selection bias is perfectly
(anti-)correlated with the shear signal.  This
means that shear selection bias
takes the form of a calibration error, with the fractional error
depending on the object
detection algorithm and on the underlying population
of galaxies.  As previously mentioned, calibration errors are difficult
to detect using, for
example, E/B decomposition of the measured shear field, which is
insensitive to a position-independent calibration error.

We organize this section as follows: in Section \ref{sec:ssbeffect}, we
develop a formalism for relating the galaxy selection probability to the
selection
bias.  In Sections \ref{sec:sta} and \ref{sec:multiscale}, we apply this
formalism to the problem of evaluating shear selection bias for galaxy
detection algorithms based
on a statistical significance threshold.  We show in Section
\ref{sec:ssbimp} that for this class of algorithms, shear selection bias
is an effect of order several per cent.
Suggestions for overcoming the shear selection bias are offered in
Section \ref{sec:correctssb}.

\subsection{\label{sec:ssbeffect}Effect on Measured Shear}

Here we analyze the effect of shear selection bias on measurements of
gravitational shear.  Following KSB95, we will use
lowercase Roman letters for vector indices and Greek letters for spin-2
tensor (i.e. traceless symmetric
matrix) indices.  To avoid confusion, we call an object with two tensor
indices a ``hypertensor''.

A shear estimator $\hat\gamma$ will generally take the form:

\beq
(\hat\gamma^+ ,
\hat\gamma^\times)
= {1\over \sum_J w_J} {\cal T}^{-1} \sum_J
w_J ( \epsilon_J^+ ,
\epsilon_J^\times
),
\label{eq:gammaest}
\eeq
where the sum is over the background galaxies in the region of sky where
the shear is to be estimated,
 $\epsilon_J$ is a measure of the shape of a galaxy, $w_J$ are the
weights for each galaxy, and $\cal T$ is a 
calibration factor.  In order to have the proper first-order behavior, we
must have:

\beq
{\cal T}\delta^\alpha{_\beta} = {1\over\langle w\rangle_d} \left<
{\partial(w\epsilon^\alpha)\over\partial\gamma^\beta} \right>_d,
\label{eq:t}
\eeq
where the $d$ subscript indicates that the average is a weighted average
taken over the detected background galaxies.  Because shears observed in
weak lensing experiments
have $\gamma\ll 1$, there is typically no need for correct $O(\gamma^2)$
and higher behavior.

In the BJ02 formalism, the
shape $\epsilon_J={\bf e}_J$, where 
${\bf e}_J$ is the galaxy's ellipticity.  In the KSB95 formalism,
$\epsilon_J$ is a shear estimator computed according
to:

\beq
\epsilon^\alpha = [P^{\gamma-1}]^\alpha{_\beta} \left( \check e^\beta -
[P^{\rm sm-1}]^\beta{_\delta}p^\delta
\right)
,
\eeq
where $\check e^\beta$ is the circular-weighted ellipticity:

\beq
(\check e^+,\check e^\times)
= {1\over \int_{-\infty}^\infty \int_{-\infty}^\infty
(x^2+y^2) W(r^2) I(x,y) \rmd x \rmd y }
\int_{-\infty}^\infty \int_{-\infty}^\infty
(x^2-y^2,2xy) W(r^2) I(x,y) \rmd x \rmd y
,
\label{eq:checke}
\eeq
$p$ is the unweighted quadrupole moment of the PSF, and the hypertensors
$P^{\rm sm}$ and $P^\gamma$ are the smear and
shear polarizabilities, respectively.
The weight function $W(r^2)$ is usually taken as a Gaussian.  The smear
polarizability is chosen to eliminate the PSF
bias (i.e. the change in measured galaxy ellipticity due to elongation of
the PSF), and the shear polarizability is
chosen so as to achieve the correct response, i.e. equation
(\ref{eq:t}) with ${\cal T}=1$:

\beq
[P^{\gamma}]^\alpha{_\beta} = {\partial \check
e^\alpha\over\partial\gamma^\beta}.
\eeq
Methods for computing $P^\gamma$ can be found in terms of Laguerre
coefficients in our Appendix \ref{sec:ksb} and in
terms of real-space variables in \citet{1998NewAR..42..137H}'s Appendix A
[these methods are based on KSB95 and \citet{1997ApJ...475...20L},
however KSB95's
shear polarizability contains algebra errors that have been corrected by
\citet{1998NewAR..42..137H}].

The effect of shear selection bias can now be analyzed.  If we have an
ensemble of background galaxies
gravitationally sheared by a small amount $\gamma$, then the shear
estimated according to equation
(\ref{eq:gammaest}) is:

\beqa
\langle\hat\gamma^\alpha\rangle && = \left< {\cal T}^{-1} \epsilon^\alpha
\right>_{wd} 
 = {1\over\cal T} { \left< \epsilon^\alpha w{\cal P} \right> \over \left<
w{\cal P} \right> }
\nonumber\\ &&
\approx{1\over{\cal T}\left< w{\cal P} \right>} \left[ 
\left.
\left< \epsilon^\alpha w{\cal P} \right>
\right|_{\gamma=0}
+ \left( \left< {\partial\epsilon^\alpha\over\partial\gamma^\beta} {w\cal
P} \right>
+ \left< \epsilon^\alpha {\cal P}{\partial{w}\over\partial\gamma^\beta}
\right>
+ \left< \epsilon^\alpha w{\partial{\cal P}\over\partial\gamma^\beta}
\right>
- { \left< \epsilon^\alpha {w\cal P} \right> \over \left< w{\cal P}
\right> } w\left< {\partial (w{\cal
P})\over\partial\gamma^\beta} \right>
\right) \gamma^\beta
\right]
,
\eeqa
where
 $\cal P$ indicates the
probability of a galaxy being detected.  (Detection is a stochastic
process due to photon counting noise.)  Assuming PSF
anisotropies and camera shear are correctly removed, the terms
proportional to $\epsilon^\alpha\cal P$ vanish.  This
yields:

\beqa
\langle\hat\gamma^\alpha\rangle && = \left< {\cal T}^{-1} \epsilon^\alpha
\right>_{wd}
\approx
{1\over{\cal T}\left< w{\cal P} \right>}
\left[
\left< {\partial\epsilon^\alpha\over\partial\gamma^\beta} w{\cal P}
\right> +
 \left< \epsilon^\alpha w{\partial{\cal P}\over\partial\gamma^\beta}
\right> +
 \left< \epsilon^\alpha {\cal P}{\partial{w}\over\partial\gamma^\beta}
\right>
\right]
\gamma^\beta
\nonumber\\ &&
=
{1\over{\cal T}} \left< 
{\partial\epsilon^\alpha\over\partial\gamma^\beta} +
\epsilon^\alpha {\partial\over\partial\gamma^\beta}\ln w +
\epsilon^\alpha {\partial\over\partial\gamma^\beta}\ln{\cal P} 
\right>_{wd}
\gamma^\beta
\equiv
{1\over{\cal T}} \left<
{\partial\epsilon^\alpha\over\partial\gamma^\beta}+\epsilon^\alpha
{\partial\over\partial\gamma^\beta}\ln w + \Xi^\alpha{_\beta} \right>_{wd}
\gamma^\beta
,
\eeqa
where we have defined the transfer hypertensor $\Xi$.  
Note that if the
estimator's response is correctly calibrated in the absence of selection
biases [i.e. equation (\ref{eq:t}) is
satisfied], then we have:

\beq
\left< {\partial\epsilon^\alpha\over\partial\gamma^\beta} +
\epsilon^\alpha {\partial\over\partial\gamma^\beta}\ln w \right>_{wd}
= {\cal T}\delta^\alpha{_\beta}
\eeq
and

\beq
\langle\hat\gamma^\alpha\rangle =\gamma^\alpha + {1\over{\cal
T}}\left<\Xi^\alpha{_\beta} \right>_{wd}
\gamma^\beta
,
\label{eq:bias}
\eeq
 in which case $\Xi$ represents
the calibration error resulting from shear selection bias.
Thus if we can determine $\Xi$, or even its average value, the shear
selection bias is overcome.  
Below, we examine the transfer hypertensor for several object detection
algorithms, and
find that $\Xi$ is strongly dependent on the details of this algorithm.

\subsection{\label{sec:sta}Significance Threshold Algorithms}

The simplest object detection algorithm is the significance threshold
algorithm.  This method convolves the
measured image intensity $I({\bmath x})$ with some detection kernel $K_1$
to
yield an convolved intensity $J$:

\beq
J({\bmath x}) = \int_{{\bf R}^2} I({\bmath x}') K_1({\bmath x}-{\bmath
x}') \rmd^2{\bmath x}'.
\label{eq:j}
\eeq
This quantity has an expectation value:

\beq
\bar J({\bmath x}) = \langle J({\bmath x}) \rangle
= \int_{{\bf R}^2} f({\bmath x}') \left[ \int_{{\bf R}^2} K_1({\bmath
x}-{\bmath
x}'') g({\bmath x}''-{\bmath x}') \rmd^2{\bmath x}'' \right]
\rmd^2{\bmath x}'
= \int_{{\bf R}^2} f({\bmath x}') K({\bmath x}-{\bmath x}') \rmd^2{\bmath
x}',
\eeq
where $f$ is the intrinsic intensity of the galaxy, $g$ is the PSF, and
$K$ is the convolution of $K_1$ and $g$.  The
two-point correlation function of the noise is simply the autocorrelation
of the kernel:

\beq
\xi({\bmath y}) =
\langle \delta J({\bmath x}) \delta J({\bmath x}+{\bmath y}) \rangle
= n \int_{{\bf R}^2} K_1({\bmath z}) K_1({\bmath y}+{\bmath
z}) \rmd^2{\bmath z},
\label{eq:xi}
\eeq
where $n$ is the noise variance per unit area.  We will require $K_1$ to
be normalized so that $\xi(0) = 1$.  Then an
object is ``detected'' if at least one point within it has $J({\bmath
x})\ge\nu_0$, where $\nu_0$ is the cutoff
significance.  Note that with our normalization, we can think of the
point ${\bmath x}$ as being $J({\bmath x})$ standard
deviations above background.

Our principal objective here is to determine the probability $\cal P$ of
the object being detected.  The definition of
``detected'', and hence the determination of $\cal P$, is a delicate
issue.  If we simply assume some intrinsic
intensity $f({\bmath x})$ peaked at ${\bmath x}=0$ and ask whether
$J(0)\ge\nu_0$, we will underestimate $\cal P$ because
$J$ is a random field and it is unlikely that $0$ will be an exact
maximum of $J$.  However, the use of $\nu=J(0)$ as a
detection statistic will serve as a first
approximation to the shear selection bias, and due to the complexity of
the problem it is unclear whether substantial
gains can be made through more detailed analytical investigation.

The probability of a galaxy being detected is given by the cumulative
Gaussian distribution function:

\beq
{\cal P} = {\cal P}(\nu\ge\nu_0) = {1\over 2}+ {1\over 2}\erf
{\bar\nu-\nu_0\over\sqrt 2}
,
\label{eq:pform}
\eeq
where $\bar\nu = \bar J(0)$ is the expected significance of the galaxy.
Defining the function:

\beq
T(\bar\nu;\nu_0) ={\partial\ln{\cal P}\over\partial\ln\bar\nu} =\frac{
{1\over\sqrt{2\pi}} \exp\left[ - {(\bar\nu-\nu_0)^2\over 2} \right]
}{
{1\over 2}+ {1\over 2}\erf {\bar\nu-\nu_0\over\sqrt 2}
}\bar\nu
,
\label{eq:tfactor}
\eeq
we can then write the transfer hypertensor as:

\beq
\Xi^\alpha{_\beta} = \epsilon^\alpha {\partial\over\partial\gamma^\beta}
\ln {\cal P}
= \epsilon^\alpha {\partial\ln{\cal P}\over\partial\ln\bar\nu}
{\partial\ln\bar\nu\over\partial\gamma^\beta}
=\epsilon^\alpha
T(\bar\nu;\nu_0) {\partial\ln\bar\nu\over\partial\gamma^\beta}
.
\label{eq:bigxi}
\eeq

We can compute the partial derivative
$\partial\ln\bar\nu/\partial\gamma^\beta$ using integration by parts:

\beq
{\partial\ln\bar\nu\over\partial\gamma^\beta} = \left. \frac{
{\partial\over\partial\gamma^\beta}
\int_{{\bf R}^2} f(S_\gamma{\bmath x}) K(-{\bmath x}) \rmd^2{\bmath x}
}{
\int_{{\bf R}^2} f(S_\gamma{\bmath x}) K(-{\bmath x}) \rmd^2{\bmath x}
} \right|_{\gamma=0}
=
\delta_\beta^{ij}
\frac{
\int_{{\bf R}^2}  x_i[ {\partial\over\partial x^j} f({\bmath x})]
K(-{\bmath
x}) \rmd^2{\bmath x}
}{
\int_{{\bf R}^2} f({\bmath x}) K(-{\bmath x}) \rmd^2{\bmath x}
}
=
-\delta_\beta^{ij}
\frac{
\int_{{\bf R}^2} f({\bmath x}) x_i{\partial\over\partial x^j} K(-{\bmath
x}) \rmd^2{\bmath x}
}{
\int_{{\bf R}^2} f({\bmath x}) K(-{\bmath x}) \rmd^2{\bmath x}
}
,
\eeq
where the mixed hypertensor $\delta_\beta^{ij}$ has components:

\beq
\delta_+ = \left(\begin{array}{cc}
1 & 0 \\ 0 & -1
 \end{array}\right)
{\rm\hskip 0.1in and\hskip 0.1in}
\delta_\times = \left(\begin{array}{cc}
0 & 1 \\ 1 & 0
 \end{array}\right)
\eeq
and we have used the tracelessness of $\delta_\beta$ to eliminate the
$\partial x_i\over\partial x^j$ term.  In the
special case where $K$ is a Gaussian, $K({\bmath x}) \propto \exp
(-{1\over
2}|{\bmath x}|^2/\sigma_K^2)$, this reduces to:

\beq
{\partial\ln\bar\nu\over\partial\gamma^\beta} = -{1\over\sigma_K^2} {
\int_{{\bf R}^2} \delta_\beta^{ij} x_i x_j f({\bmath x}) \rme^{-|{\bmath
x}|^2/2\sigma_K^2} d^2{\bmath x}
\over
\int_{{\bf R}^2} f({\bf x}) \rme^{-|{\bmath x}|^2/2\sigma_K^2} d^2{\bmath
x}
}
=
-{1\over\sigma_K^2} {
\int_{{\bf R}^2} |{\bmath x}|^2 f({\bmath x}) \rme^{-|{\bmath
x}|^2/2\sigma_K^2}
d^2{\bmath x}
\over
\int_{{\bf R}^2} f({\bf x}) \rme^{-|{\bmath x}|^2/2\sigma_K^2} d^2{\bmath
x}
}
\check e_\beta
.
\label{eq:gaussfilter}
\eeq
For a single detection kernel $K_1$, equation
(\ref{eq:gaussfilter}) describes the transfer hypertensor.  However, the
single-scale detection algorithms considered here
are not optimal for detecting objects substantially larger or smaller
than the size of the detection kernel.  For this reason, most authors
have used multiple detection
kernels, which we consider next.

\subsection{\label{sec:multiscale}Multiscale Detection Algorithms}

The commonly used KSB95 detection algorithm (see, e.g. KSB95,
\citealt{1998NewAR..42..137H}) detects objects by running a
significance
threshold algorithm on several
smoothing scales, i.e. with several functions $K_1$.  We will analyze
here only the case of a Gaussian PSF of standard
deviation $\sigma_g$ and a Gaussian smoothing filter of adjustable
standard deviation $\sigma_d$.  Then we have:

\beqa
  g({\bmath x}) = && {1\over 2\pi\sigma_g^2} \rme^{-|{\bmath
x}|^2/2\sigma_g^2}
\nonumber\\
K_1({\bmath x}) = && {1\over\sigma_d\sqrt{\pi n}} \rme^{-|{\bmath
x}|^2/2\sigma_d^2} \nonumber\\
  K({\bmath x}) = && {\sigma_d\over\sigma_K^2 \sqrt{\pi n}}
\rme^{-|{\bmath
x}|^2/2\sigma_K^2},
\eeqa
where $\sigma_K^2=\sigma_d^2+\sigma_g^2$.
An object is detected if it is detected with any of these filters.  We
must therefore evaluate equation
(\ref{eq:gaussfilter}) at the most significant scale, i.e. where
$\bar\nu$ is maximized with respect to $\sigma_d$.  We
determine $\bar\nu$ by integration:

\beq
\bar\nu = \bar J(0) = {\sigma_d\over\sigma_K^2 \sqrt{\pi n}} \int_{{\bf
R}^2} f({\bmath x}) 
\rme^{-|{\bmath x}|^2/2\sigma_K^2} \rmd^2{\bmath x}.
\eeq
The maximum significance occurs where
$\partial\bar\nu/\partial\sigma_d=0$, that is where:

\beq
\frac{
\int_{{\bf R}^2} I({\bmath x}) |{\bmath x}|^2 \rme^{-|{\bmath
x}|^2/2\sigma_K^2}
\rmd^2{\bmath x}
}{
\int_{{\bf R}^2} I({\bmath x}) \rme^{-|{\bmath x}|^2/2\sigma_K^2}
\rmd^2{\bmath x}
} =
\sigma_K^2 \left( 2-{\sigma_K^2\over\sigma_d^2}\right)
=
\sigma_K^2 \left( 1-{\sigma_g^2\over\sigma_d^2}\right)
.
\label{eq:match}
\eeq
(Remember that $\sigma_K$ depends implicitly on $\sigma_d$.)  In the
special case of a circular Gaussian object $f\propto\exp(-{1\over
2}|{\bmath
x}|^2/\sigma_f^2)$, the
left-hand side of equation (\ref{eq:match}) becomes
$2/(\sigma_K^{-2}+\sigma_f^{-2})$; this gives a solution to equation
(\ref{eq:match}) of:

\beqa
{2\over\sigma_K^{-2}+\sigma_f^{-2}} = && \sigma_K^2 \left(
1-{\sigma_g^2\over\sigma_d^2} \right) \nonumber\\ 
{2\over(\sigma_d^2+\sigma_g^2)^{-1}+\sigma_f^{-2}} = &&
(\sigma_g^2+\sigma_d^2) \left( 1-{\sigma_g^2\over\sigma_d^2}
\right) \nonumber\\  
\sigma_d^2 = && \sigma_g^2+\sigma_f^2.
\eeqa
 (We ignore the imaginary solution $\sigma_d^2=-\sigma_g^2$.)  This means
 that if the galaxy is a circular Gaussian, the detection algorithm will
return maximum
significance with a filter matched to the object size.

 Whether or not the galaxy is Gaussian, we can use equation
 (\ref{eq:match}) to convert the logarithmic derivative expression,
equation (\ref{eq:gaussfilter}), into:

\beq
{\partial\ln\bar\nu\over\partial\gamma^\beta}
= - \left( 1-{\sigma_g^2\over\sigma_d^2}\right) \check e_\beta
\approx
 - \left( 1-{\sigma_g^2\over\sigma_d^2}\right) \epsilon_\beta,
\eeq
since for near-circular Gaussian objects, $\epsilon\approx\check{\bf
e}$.  (This follows because the smear polarizability $P^\gamma$ is
approximately the identity matrix in this case, as can be verified through
direct computation using the formulas in Appendix \ref{sec:ksb}.)  The
transfer hypertensor is then [see equation (\ref{eq:bigxi})]:

\beq
\Xi^\alpha{_\beta} = - T(\bar\nu;\nu_0)  \left(
1-{\sigma_g^2\over\sigma_d^2}\right)
\epsilon^\alpha \epsilon_\beta.
\label{eq:findxi}
\eeq

\begin{figure}
\includegraphics[angle=-90, width=4in]{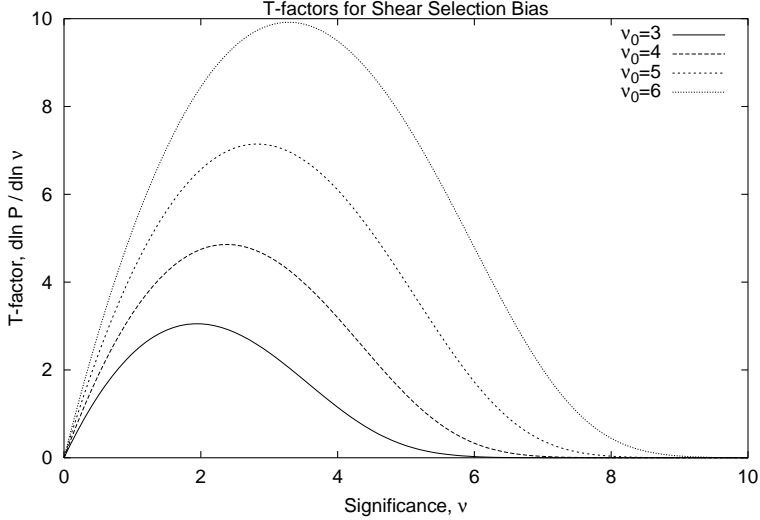}
\caption{\label{fig:figt}The $T$-factor $T(\bar\nu;\nu_0)=d\ln{\cal
P}/d\ln\bar\nu$.  This factor represents the (logarithmic) increase in
probability of detection $\cal
P$ if we increase the significance of the galaxy.  The shear selection
bias increases in proportion to an appropriate average of $T$ [see
equation (\ref{eq:bigxi})].  Note
that $T\rightarrow 0$ as $\bar\nu\rightarrow\infty$; this means that
there is no shear selection bias for objects far above threshold, as
expected.
}
\end{figure}

\subsection{\label{sec:ssbimp}How Important is Shear Selection Bias?}

We can evaluate, at least at the order-of-magnitude level, the importance
of the shear selection bias.  Since $\sigma_d$
is matched to the size of the object,
the quantity $1-\sigma_g^2/\sigma_d^2$ will be near unity for
well-resolved galaxies, decreasing to zero for point-like objects.  (That
$\Xi=0$ for pointlike objects is
unsurprising because a point source remains a point source when
sheared.)  The $T$-factor is shown in Fig. \ref{fig:figt}.  
The weighted ellipticities $\check{\bf e}$
are half of the adaptive ellipticities [equation (\ref{eq:edef})] for
near-circular Gaussian objects with a matched Gaussian weight
function.  Thus the $\check{\bf e}$
circular ellipticity has shape variance of approximately
$V\approx(0.32/2)^2\approx 0.026$, where 0.32 is the
RMS ellipticity of
galaxies \citep{2001astro.ph..8013M}.
(This parameter will depend somewhat on the
sample examined, so the value 0.32 should be
thought of as a rough guide rather than a law of nature.)
It thus appears that the shear
selection bias will be of order several percent for
well resolved galaxies, 
falling to zero in the limit of a poorly resolved object.

A more quantitative approach to the problem is possible in the special
case where the galaxy significance distribution is a power law and the
shear estimator is
constructed from an unweighted distribution of galaxies
(i.e. $w$=constant).  We suppose that the
galaxy counts scale as $N\propto\bar\nu^{-\lambda}$.
 The exponent $\lambda$ is determined by noting that $\bar\nu$ is a
signal-to-noise ratio and hence is proportional to the galaxy's total
flux $F$ divided by the
square
root of its post-PSF-convolved area $A$; the number-magnitude relation is
roughly $N\propto F^{-1}$.  If the galaxies are poorly resolved, $A$ is
determined
by the PSF and hence is constant, in which case $\lambda=1$.  For
well-resolved, nearby galaxies (redshift $z\ll 1$) we can assume a
Euclidean universe, in which the usual
inverse square
laws $F\propto A\propto 1/r^2$ apply; this gives $F/\sqrt{A}\propto
1/r\propto N^{-1/3}$, since the number of objects scales as $r^3$.  This
gives $\lambda=3$.  Thus we
expect that $\lambda$ will probably lie in the range $1\le\lambda\le 3$,
with the distant galaxies of interest to cosmic shear measurements likely
having $\lambda$ closer
to 1.

We next evaluate the average of equation (\ref{eq:findxi}).  We define a
shear selection bias resolution factor ${\cal
R}=1-\sigma_g^2/\sigma_d^2$, which is 1 for
well-resolved and 0 for poorly-resolved objects.  We also use that
$dN/d\bar\nu\propto\bar\nu^{-1-\lambda}$.  The average transfer
hypertensor is:

\beqa
\langle \Xi^\alpha{_\beta} \rangle_{wd} &&
=- \frac{
\int_0^\infty \bar\nu^{-1-\lambda} {\cal PR}
T(\bar\nu;\nu_0)  \epsilon^\alpha\epsilon_\beta \rmd\bar\nu
}{
\int_0^\infty \bar\nu^{-1-\lambda} {\cal P} \rmd\bar\nu
}
=
-{\cal R}_{eff} V\delta^\alpha{_\beta}
\frac{
\int_0^\infty \bar\nu^{-1-\lambda} {\cal P} T(\bar\nu;\nu_0) \rmd\bar\nu
}{
\int_0^\infty \bar\nu^{-1-\lambda} {\cal P} \rmd\bar\nu
}
\nonumber\\ &&
=
-{\cal R}_{eff} V\delta^\alpha{_\beta}
\frac{
\int_0^\infty \bar\nu^{-\lambda} {\partial{\cal P}\over\partial\bar\nu}
\rmd\bar\nu
}{
\int_0^\infty \bar\nu^{-1-\lambda} {\cal P} \rmd\bar\nu
}
,
\label{eq:inf}
\eeqa
where ${\cal R}_{eff}$ is a weighted (by $T$) average of the shear
selection bias resolution factor $\cal R$ over the detected
galaxies: ${\cal R}_{eff}=\langle {\cal
R}T\rangle_d/\langle T\rangle_d$.  Since $T$ is a rapidly decreasing
function of $\bar\nu$ (see Fig. \ref{fig:figt}), ${\cal R}_{eff}$ should
be evaluated for galaxies
near the detection threshold.

At this point we note that the integrals in the numerator and denominator
of equation (\ref{eq:inf}) diverge at the $\bar\nu\rightarrow 0$ limit
because ${\cal
P}(0)>0$; see equation (\ref{eq:pform}).  However, this divergence must
 be unphysical since real source galaxy catalogs do
not contain an infinity of faint objects.  [The origin of the problem is
that in deriving equation (\ref{eq:pform}), we neglected the fact that
objects overlap with each
other.]  We will thus remove this divergence by supposing that at very
small $\bar\nu$, the detection
probability $\cal P$ vanishes sufficiently rapidly for the integrals in
equation (\ref{eq:inf}) to converge; this yields (using integration by
parts):

\beq
\langle \Xi^\alpha{_\beta} \rangle_{wd} =
-{\cal R}_{eff} V\delta^\alpha{_\beta}
\frac{
\int_0^\infty \bar\nu^{-\lambda} {\partial{\cal P}\over\partial\bar\nu}
d\bar\nu
}{
\int_0^\infty \bar\nu^{-1-\lambda} {\cal P} d\bar\nu
}
=
{\cal R}_{eff} V\delta^\alpha{_\beta}
\frac{
\int_0^\infty {\cal P}
{\partial\over\partial\bar\nu}(\bar\nu^{-\lambda}) d\bar\nu
}{
\int_0^\infty \bar\nu^{-1-\lambda} {\cal P} d\bar\nu
}
=
-\lambda {\cal R}_{eff} V\delta^\alpha{_\beta}
\label{eq:xilambda}
\eeq
Thus for cosmic shear measurements where galaxies are selected via a
statistical significance cut, where $V=0.026$ and
$\lambda\approx 1$, the shear selection bias causes 
an underestimate of the shear by 2.6${\cal R}_{eff}$ per cent.  (It
vanishes for the poorly resolved galaxies with
$R_{eff}\approx 0$.)
This factor would be increased to roughly 8 per cent in the
case of $\lambda=3$, $R_{eff}=1$, i.e. of
low-redshift source galaxies for which the Euclidean source counts are
more nearly valid and
the near-threshold objects are well-resolved.  
Note that this result is independent of the significance threshold, 
at least when a power-law form for the
galaxy
count is assumed.  Indeed, because it does not depend on the particular
form for ${\cal P}$, this result is still valid if
only those source galaxies are selected that appear in multiple images of
the same region of sky (this is the case in most current lensing surveys).
  It is also important to understand that shear selection bias is small
not because the shear is small, but because the background 
galaxies are so nearly circular. If the 
galaxies are more elliptical in deeper surveys than in SDSS used here 
the bias will be larger. Note also that
the mean RMS ellipticity, $\lambda$
and $R_{eff}$ are all 
likely to be redshift dependent and so will be the bias.

\subsection{\label{sec:correctssb}Correcting for Shear Selection Bias}

While the above analysis serves to illustrate the order of magnitude of
the shear selection bias problem, it may not be adequate to provide an
analytical correction due to
the large number of assumptions and approximations made.  There are also
contributions to the selection bias other than significance
thresholds; for example, it is
generally necessary to remove overlapping objects from the background
galaxy catalog, but the overlap probability is a function of galaxy
shape.  There are four major
approaches that could be taken to overcome shear selection bias:

\newcounter{Lcount1}
\begin{list}{\arabic{Lcount1}.}{\usecounter{Lcount1}}
\item\label{it:ig} The bias could simply be ignored.
\item\label{it:an} The galaxy detection algorithm could be designed so as
to eliminate, or at least substantially reduce, the bias.
\item\label{it:polar} The transfer hypertensor $\Xi$ could be computed and
used as a correction to the shear polarizability or resolution factor.
\item\label{it:sim} The bias could be determined from simulations.
\end{list}

Of these, Method \ref{it:ig} is appropriate when calibration errors of
order several per cent are unimportant, for example in typical galaxy-galaxy
lensing studies where
statistical
error dominates.  However, it is clearly inapplicable to future cosmic
shear surveys that aim for much higher accuracy.  Method \ref{it:an}
requires a detection statistic
that does not suffer from shear selection bias, i.e. one that is
independent of the shear.  Examples would be the galaxy's magnitude in a
given band and the
PSF-deconvolved area.  Unfortunately, these are not truly independent of
the shear when noise is taken into account, and the ``overlapping object''
effect described above
is not addressed.  These effects are difficult to compute analytically,
hence it is not clear whether we can verify that the residual errors in
Method \ref{it:an} are
``small enough'' without resorting to simulation.  

Method \ref{it:polar} requires that we compute $E_\beta =
{\partial\over\partial\gamma^\beta}\ln{\cal P}$ for each
galaxy in our survey, then find $\Xi^\alpha{_\beta}=\epsilon^\alpha
E_\beta$, and compute the weighted average $\langle
\Xi^\alpha{_\beta}\rangle_{wd}$ to determine the calibration error due to
shear selection bias.  The problem is thus one
of finding $E_\beta$; unfortunately, $E_\beta$ can only be determined from
the intrinsic intensity $f$ of the galaxy,
not the observed image.
Thus correction of cosmic shear measurements based on determination of
$\Xi$ would require a careful
analysis to ensure that noise and PSF effects do not invalidate the
calculation.  Method \ref{it:sim}, the simulation,
is straightforward albeit time-consuming. Its advantage is that it can
accommodate arbitrarily complicated detection
algorithms.  The main weakness is that the underlying distribution of
galaxies in the simulation may not be
correct; however, the shear selection bias is only an effect of order
several
per cent anyway, so determination of the bias to many
significant figures is unnecessary.

\section{Conclusion}
\label{sec:4}

In this paper we have addressed several possible biases present in the 
data reduction of cosmic shear measurements. We focus on effects that 
affect the overall normalization only and cannot be detected using the 
usual methods of finding systematics, such as the tests on stars and 
E/B decomposition. We pay particular attention 
to the deviations from the elliptical Gaussian profile of the galaxy 
or the PSF, which is often the starting assumption of the data reduction 
methods. 
The results in Figs. \ref{fig:sim0}--\ref{fig:sim5} show that while 
for a Gaussian PSF most of the methods are successful in reconstructing
the 
shear, the case of a more realistic non-Gaussian PSF is significantly 
more problematic except for the re-Gaussianization 
method introduced in this paper. This suggests that in existing surveys 
the shear calibration may be inaccurate at the 10 per cent level. 
Some methods, such as current implementation 
of BJ02 used in \citet{2002astro.ph.10604J} and
\citet{2001astro.ph..8013M},
 are particularly susceptible to this and can underestimate the linear 
amplitude of fluctuations by 10 per cent, 
while those based on KSB95 (e.g. \citealt{2000MNRAS.318..625B},
\citealt{2000A&A...358...30V},
\citealt{2001ApJ...552L..85R},
\citealt{2002ApJ...572...55H}, \citealt{2002astro.ph.10213B})  have a
negative bias of up to 5 per cent.
It is interesting to note that \citet{2002astro.ph.10604J} 
find a significantly (10--20 per cent) lower
value for $\sigma_8$ than other surveys. Calibrations errors of the 
type discussed here can explain part of the difference,
although other factors such 
as statistical errors, 
uncertainties in the redshift distribution of background galaxies, 
nonlinear evolution corrections of dark matter power spectrum
and residual systematics (which are known to be present because their 
effect on B-type correlation is seen in all the surveys that have 
looked for it) also play a role. At the moment, residual systematics
due to the PSF correction can be as great as 10 per cent in $\sigma_8$ and 
are comparable to the statistical errors in current samples.

There are other biases that one must worry about when accuracy on the
order of one per cent
is required. One such effect 
is shear selection bias which causes the galaxy selection to be 
correlated with the underlying shear and with the galaxy flux.
This error is of order a few percent, but can 
reach up to roughly 8 per cent in some limits.  
Since the shear selection bias preferentially 
selects galaxies with shear orthogonal to cosmic shear the bias
translates into an 
overall underestimate of the amplitude of fluctuations.  
Moreover, since for a flux limited survey the 
smaller and fainter
galaxies are at a higher redshift the error in the amplitude
may be correlated with redshift and can give an incorrect growth 
factor if not properly accounted for. Another effect that can 
create this is if the mean RMS ellipticity of the galaxies 
changes with redshift.

Looking further ahead, weak lensing has the potential to break many 
of the degeneracies present in the cosmological parameters even after 
the cosmic microwave background has given us all of its information,
especially 
if one can use the redshift information on background galaxies 
to extract the information on the growth factor
\citep{1999ApJ...522L..21H}. 
It is an
especially powerful probe of dark energy and massive neutrinos,
since they both affect the growth factor at low redshifts probed by  
weak lensing. While the statistical precision of some of the 
proposed experiments such as LSST is impressive, the systematic 
effects must also be controlled at 1 per cent level. Current 
reduction methods are not sufficient for this purpose, but improved 
methods such as our re-Gaussianization method (which can be 
further generalized to include higher order terms) should be up to the 
task. Similarly, shear selection bias and other biases 
should be carefully simulated 
to estimate their importance as a function of galaxy detection 
significance, PSF etc.  One notes that realistic simulations would allow 
us to determine both the PSF dilution calibration and the shear selection 
bias, and indeed shear selection bias is implicitly taken into account in 
simulated observations such as those of \cite{2000ApJ...537..555K}, 
\cite{2001A&A...366..717E}, and \cite{2001MNRAS.325.1065B}.  Future 
experiments will require that this line of work be extended, for example 
to include measured PSFs from the telescope and the correct distribution 
of galaxy shapes and fluxes as a function of redshift.

It is clear that the quality of the data of the upcoming measurements 
will exceed the current generation of the reduction methods. 
However, this is not a cause for alarm, since none of the problems 
discussed here is fundamental in nature. We have proposed 
solutions for some of the problems, while others will require 
more detailed modeling using realistic simulations of the galaxies 
and their image processing through the atmosphere (if applicable),
telescope, detector, and software.
Given enough attention to these details weak lensing should fulfill 
its promise in the era of high precision cosmology. 

\section*{Acknowledgments}

The authors acknowledge useful discussions with Gary Bernstein,
Mike Jarvis, Robert
Lupton, 
Michael Strauss, Joseph Hennawi, and Nikhil Padmanabhan.
CH is supported through the NASA Graduate Student Researchers Program,
grant NASA GSRP-02-OSS-079.
This work was supported by NASA NAG5-1993 and NSF CAREER. US is a fellow
of 
David and Lucille Packard Foundation and Alfred P. Sloan Foundation.

\appendix

\section{Convolution with Elliptical Laguerre Expansion}
\label{sec:conv}

Here we examine the problem of convolving two functions $f$ and $g$ given
in terms of their Laguerre expansions:

\beq
\left\{
\begin{array}{c}
f({\bmath x}) = \sum_{p,q} a_{pq} \psi_{pq}({\bmath x};{\bf M}_f)
\\
g({\bmath x}) = \sum_{p,q} b_{pq} \psi_{pq}({\bmath x};{\bf M}_g)
\end{array}\right.
.
\eeq
The convolution is:

\beq
h({\bmath x}) = \int_{{\bf R}^2} f({\bmath x}') g({\bmath x}-{\bmath
x}') d^2{\bmath x}'
= \sum_{p,q} c_{pq} \psi_{pq}({\bmath x};{\bf 
M}_h)
;
\label{eq:h1}
\eeq
we will consider here the case where ${\bf M}_h={\bf M}_f+{\bf M}_g$.
This problem is easiest in Fourier space using the convolution
theorem $\tilde h({\bmath k}) = 2\pi\tilde f({\bmath k})\tilde g({\bmath
k})$; using equation (\ref{eq:fourier}) and
writing the result in terms of $\hat\kappa = {\bf M}_h^{1/2}{\bmath k}$
yields:

\beq
\sum_{pq} {\rmi^N\over\sqrt{\pi\cdot p!q!}} c_{pq}
\Lambda_{pq}(\hat\kappa)
= 2
\sum_{p'q',p''q''} {\rmi^{N'+N''}\over\sqrt{p'!q'!p''!q''!}} a_{p'q'}
b_{p''q''}
\Lambda_{p'q'}({\bf M}_f^{1/2}{\bf M}_h^{-1/2}\hat\kappa)
\Lambda_{p''q''}({\bf M}_g^{1/2}{\bf M}_h^{-1/2}\hat\kappa).
\label{eq:csum}
\eeq
We can simplify this by defining the matrices ${\bf A}^{(f)}={\bf
M}_f^{1/2}{\bf M}_h^{-1/2}$ and similarly ${\bf
A}^{(g)}$.  Then
we use the result that any $2\times 2$ matrix, whether symmetric or not,
can be written as a product of a rotation
matrix, a diagonal matrix, and another rotation matrix (we will call this
the RDR decomposition):

\beq
{\bf A} =
\left(
\begin{array}{cc} \cos\alpha & -\sin\alpha \\ \sin\alpha & \cos\alpha
\end{array}
\right)
\left(
\begin{array}{cc} \bar C & 0 \\ 0 & \bar D \end{array}
\right)
\left(
\begin{array}{cc} \cos\chi & -\sin\chi \\ \sin\chi & \cos\chi \end{array}
\right)
\equiv
{\bf R}_\alpha
\left(
\begin{array}{cc} \bar C & 0 \\ 0 & \bar D \end{array}
\right)
{\bf R}_\chi
,
\label{eq:rdr}
\eeq
where ${\bf R}_\theta$ represents a rotation by angle $\theta$, and $\bar
C$, $\bar D$, and the rotation angles
$\alpha$ and $\chi$ can found explicitly from the rectangular-to-polar 
conversions:

\beq
\left\{ \begin{array}{l}
A_{11} + A_{22} = (\bar C+\bar D)\cos (\alpha+\chi) \\
A_{21} - A_{12} = (\bar C+\bar D)\sin (\alpha+\chi) \\
A_{11} - A_{22} = (\bar C-\bar D)\cos (\alpha-\chi) \\
A_{21} + A_{12} = (\bar C-\bar D)\sin (\alpha-\chi)
\end{array} \right.
\label{eq:rdrpolar}
.\eeq
This decomposition is unique up to several discrete
degeneracies: (1) interchange of $\bar C$ and $\bar D$ with phase
shift of $\alpha$ by $+\pi/2$ and
$\chi$ by $-\pi/2$; (2) sign change of $\bar C$ and $\bar D$ with phase
shift of $\alpha$ and $\chi$ together by
$\pi/2$; (3) phase shift of $\alpha$ and $\chi$ together by $\pi$; and
(4) phase shift of $\alpha$ and $\chi$
independently by multiples of $2\pi$.  (In the special case that $\bar
C=\pm\bar D$, there is an additional continuous 
degeneracy
in
that only $\alpha\pm\chi$, but not either angle individually, is
determined.)  If we decompose ${\bf A}^{(f)}$ and ${\bf
A}^{(g)}$, then the condition
${\bf M}_h={\bf M}_f+{\bf M}_g$ yields that ${\bf A}^{(f)T}{\bf
A}^{(f)}+{\bf A}^{(g)T}{\bf A}^{(g)}$ is the
identity; this can be expressed
in
matrix form as:

\beqa
&& \left(\begin{array}{cc}
\bar C_f^2 \cos^2\chi_f + \bar D_f^2 \sin^2\chi_f & (\bar D_f^2-\bar
C_f^2)\sin\chi_f\cos\chi_f \\
(\bar D_f^2-\bar C_f^2)\sin\chi_f\cos\chi_f & \bar C_f^2 \sin^2\chi_f +
\bar D_f^2 \cos^2\chi_f
\end{array}\right)
\nonumber\\ &&
+
\left(\begin{array}{cc}
\bar C_g^2 \cos^2\chi_g + \bar D_g^2 \sin^2\chi_g & (\bar D_g^2-\bar
C_g^2)\sin\chi_g\cos\chi_g \\
(\bar D_g^2-\bar C_g^2)\sin\chi_g\cos\chi_g & \bar C_g^2 \sin^2\chi_g +
\bar D_g^2 \cos^2\chi_g
\end{array}\right)
=
\left(\begin{array}{cc}
1 & 0 \\ 0 & 1
\end{array}\right)
.
\eeqa
The trace of this equation yields $\bar C_f^2+\bar C_g^2+\bar D_f^2+\bar
D_g^2=2$; the traceless-symmetric part yields:

\beq
(\bar C_f^2-\bar D_f^2)\left( \begin{array}{c} \cos 2\chi_f \\ -\sin
2\chi_f \end{array}\right)
+
(\bar C_g^2-\bar D_g^2)\left( \begin{array}{c} \cos 2\chi_g \\ -\sin
2\chi_g \end{array}\right)
=0,
\eeq
which implies that either $\bar C_f^2=\bar D_f^2$ and $\bar C_g^2=\bar
D_g^2$, or that $\chi_f$ and $\chi_g$ differ by a
multiple of $\pi/2$; we may, however, assume the latter case, since in
the former case the aforementioned continuous
degeneracy allows us to choose $\chi_f=\chi_g$.  Indeed, the discrete
degeneracy allows us to set
$\chi_f=\chi_g(\equiv\chi)$.  From
this we derive $\bar C_f^2+\bar C_g^2=\bar D_f^2+\bar D_g^2=1$.  Finally,
the positive determinant of ${\bf A}^{(f)}$
and ${\bf A}^{(g)}$ implies that $\bar C_f\bar D_f$ and $\bar C_g\bar
D_g$ are positive; the discrete degeneracy then
allows us to choose $\bar C_f$, $\bar C_g$, $\bar D_f$, and $\bar D_g$
positive.

Explicit expressions for $\alpha_f$, $\alpha_g$, $\chi$, $\bar C_f$,
$\bar C_g$, $\bar D_f$, and $\bar D_g$ in terms of ${\bf M}_{f,g,h}$ may
be obtained by directly finding the square roots of these matrices,
thereby determining ${\bf A}^{(f,g)}$, and then substituting these
matrices into
equation (\ref{eq:rdrpolar}).  In order to construct the ${\bf A}$'s, it
is necessary to determine the square root of a symmetric matrix.  
This is given by:

\beq
{\bf M}^{1/2} = {1\over\sqrt{M_{xx}+M_{yy}+2\sqrt{D}}} \left(
\begin{array}{cc}
M_{xx}+\sqrt{D} & M_{xy} \\
M_{xy} & M_{yy}+\sqrt{D}
\end{array}
\right)
,
\eeq
where $D=M_{xx}M_{yy}-M_{xy}^2$.

With this aside completed, we return to the determination of
$c_{pq}$.  Substituting the decomposition [equation
(\ref{eq:rdr})] into equation (\ref{eq:csum}), and setting $(U,V)={\bf
R}_{-\chi}\hat\kappa$ yields:

\beq
\sum_{pq} {\rmi^N\over\sqrt{\pi\cdot p!q!}} c_{pq} \rme^{\rmi m\chi}
\Lambda_{pq}(U,V)
= 2
\sum_{p'q',p''q''} {\rmi^{N'+N''}\over\sqrt{p'!q'!p''!q''!}} a_{p'q'}
b_{p''q''}
\rme^{-im'\alpha_f} \rme^{-im''\alpha_g}
\Lambda_{p'q'}(\bar C_fU,\bar D_fV)
\Lambda_{p''q''}(\bar C_gU,\bar D_gV),
\label{eq:csum2}
\eeq
where we have used the result that $\Lambda_{pq}({\bf R}_\theta{\bmath
k})=\rme^{-\rmi m\theta}\Lambda_{pq}({\bmath k})$.  Then
equation (\ref{eq:csum2}) is a linear equation for the $c_{pq}$, thus:

\beq
c_{pq} = 
2\sqrt\pi\cdot
\rme^{-\rmi m\chi} \sum_{p'q',p''q''}
Z^{p'q',p''q''}_{pq}(\bar C_f,\bar D_f)
a_{p'q'} b_{p''q''} \rme^{-\rmi m'\alpha_f} \rme^{-\rmi m''\alpha_g},
\eeq
where the coefficients $Z^{p'q',p''q''}_{pq}(\bar C_f,\bar D_f)$ satisfy:

\beq
\sum_{pq} {\rmi^N\over\sqrt{p!q!}} Z^{p'q',p''q''}_{pq}(\bar C_f,\bar D_f)
 \Lambda_{pq}(U,V)
=
{\rmi^{N'+N''}\over\sqrt{p'!q'!p''!q''!}}
\Lambda_{p'q'}(\bar C_fU,\bar D_fV)
\Lambda_{p''q''}(\bar C_gU,\bar D_gV).
\label{eq:zdef}
\eeq
The $Z^{p'q',p''q''}_{pq}(\bar C_f,\bar D_f)$ can be calculated by noting
that the right hand side of equation
(\ref{eq:zdef}) is a polynomial of order $p'+q'+p''+q''$ in the two
variables $U$ and $V$.  Since the
$\Lambda_{pq}(U,V)$
with $p+q\le N$ form a basis for the $(N+1)(N+2)/2$ dimensional vector
space of polynomials in $U$ and $V$ of degree
$\le N$, the sum on the left hand side of equation (\ref{eq:zdef}) may be
cut off at $p+q\le N$.  Equating coefficients
of $U^mV^n$ on both sides then results in a
finite system of linear equations for $Z^{p'q',p''q''}_{pq}(\bar C_f,\bar
D_f)$, which is easily solved using standard
linear algebra techniques.  An alternative route is to exploit the
orthonormality relations among the Laguerre basis
functions to convert equation (\ref{eq:zdef}) into:

\beq
Z^{p'q',p''q''}_{pq}(\bar C_f,\bar D_f) =
{\rmi^{N'+N''-N}\over\sqrt{p!q!p'!q'!p''!q''!}}
 \frac{
\int_{{\bf R}^2} \Lambda_{pq}^\ast(U,V)\Lambda_{p'q'}(\bar C_fU,\bar D_fV)
\Lambda_{p''q''}(\bar C_gU,\bar D_gV) \rme^{-(U^2+V^2)} \rmd U \rmd V
}{
\int_{{\bf R}^2} e^{-(U^2+V^2)} \rmd U \rmd V
};
\label{eq:wick}
\eeq
this ratio of Gaussian integrals can be computed using Wick's theorem
with the Feynman rules $\langle U \rangle =
\langle V \rangle = \langle UV \rangle = 0$ and $\langle U^2\rangle =
\langle V^2\rangle = \frac{1}{2}$.

The Gaussian-integral method allows us to construct an explicit formula
for $Z^{p'q',p''q''}_{pq}(\bar C_f,\bar D_f)$
in terms of finite sums.  To do this, we expand the $\Lambda$:

\beq
\Lambda_{pq}(u,v) = \sum_{\mu,\nu\ge 0} X_{pq\mu\nu}u^\mu v^\nu,
\eeq
where:

\beq
 X_{pq\mu\nu} = 
\left\{ \begin{array}{lcl}
\rmi^\nu \sum_t (-1)^{q-t} \frac{p!q!}{
s!(\mu-t)!(p-s-\mu+t)!t!(q-s-t)!
} & & s\ge 0,\; s{\rm ~an~integer} \\
0 & & {\rm otherwise}
\end{array} \right.,
\label{eq:sumt}
\eeq
and $s=(p+q-\mu-\nu)/2$.  The sum in equation (\ref{eq:sumt}) is over
all values of $t$ such that $s$, $\mu-t$, $t$, and $q-s-t$ are
all non-negative.  The ``selection rules'' for $X$ are as
follows: $X_{pq\mu\nu}\neq 0$ only if $\mu+\nu$ is equal to $p+q$, or
less than $p+q$ by an even integer.  Having
constructed $X$, we evaluate the integral in equation (\ref{eq:wick}) as:

\beqa
Z^{p'q',p''q''}_{pq}(\bar C_f,\bar D_f) = &&
{\rmi^{N'+N''-N}\over\sqrt{p!q!p'!q'!p''!q''!}}
\sum_{\hat\mu,\hat\nu\ge 0}
\frac{ (2\hat\mu-1)!! (2\hat\nu-1)!! }{2^{\hat\mu+\hat\nu}}
\nonumber\\ && \times
\sum_{\mu+\mu'+\mu''= 2\hat\mu} 
\sum_{\nu+\nu'+\nu''= 2\hat\nu} 
X_{pq\mu\nu}^\ast X_{p'q'\mu'\nu'} X_{p''q''\mu''\nu''}
\bar C_f^{\mu'} \bar D_f^{\nu'} (1-\bar C_f)^{\mu''} (1-\bar D_f)^{\nu''}
.\label{eq:zsum}
\eeqa

We have computed several of the lower-order (up to degree 6) $Z$
coefficients for the simulations in this paper; it is found that for
these moments, it is practical to
perform either the linear system solution [equation (\ref{eq:zdef})] or
the summation method [equation (\ref{eq:zsum})] using symbolic
manipulation software, and then
write ``hard-wired'' C code to numerically evaluate the resulting
expressions.  Because the explicit polynomial form of
the $\Lambda$ functions [see equation (\ref{eq:sumt}) contains
many positive and negative terms for large $p$ and $q$, it is likely that
round-off errors would be an issue for our
computational approach if we were to consider moments 
of sufficiently high order.  It is unlikely that moments of very high
order would be useful in weak lensing
studies; nevertheless, if this turned out to be the case, or if 
the high-order $Z$ coefficients were required for another application, it
would be necessary to develop a better
algorithm (likely involving a recursion relation) for computing them.

\section{Linear PSF Correction}
\label{sec:psfcorr}

We can derive a relation between ${\bf e}^{(f)}$ and ${\bf e}^{(I)}$,
valid to first order in the departure of the galaxy
and PSF from Gaussianity as follows.  Suppose that the
adaptive covariance of the intrinsic galaxy image is ${\bf M}_f$.  Then
the observed image $I({\bmath x})$ is the
convolution of $f$ with the PSF $g$, and it can thus be expressed as a
Laguerre expansion with covariance
${\bf M}_h={\bf M}_f+{\bf M}_g$.  Using the convolution results of
Appendix \ref{sec:conv}, we find:

\beq
I({\bmath x}) = \sum_{pq} b^{(I)}_{pq} \psi_{pq}({\bmath x};{\bf M}_I)
= \sum_{pq} c_{pq} \psi_{pq}({\bmath x};{\bf M}_h),
\eeq
where:

\beq
c_{pq} = 2\sqrt\pi \rme^{-\rmi m\chi} \sum_{p'q',p''q''}
Z^{p'q',p''q''}_{pq}(\bar C_f,\bar D_f) \rme^{-\rmi m'\alpha_f}
\rme^{-\rmi m''\alpha_g}
b^{(f)}_{p'q'} b^{(g)}_{p''q''},
\label{eq:oid1}
\eeq
and $b^{(f)}_{p'q'}$ and $b^{(g)}_{p''q''}$ are the Laguerre coefficients
of $f$ and $g$ around ${\bf M}_f$ and ${\bf
M}_g$, respectively.  (The rotation angles $\chi$, $\alpha_f$, and
$\alpha_g$ and diagonal elements $\bar C_f$ and $\bar
D_f$ can be calculated using the approximation ${\bf M}_f\approx {\bf
M}_I-{\bf M}_g$ since corrections to this result
will appear as second-order corrections to the final moments.)  Since we
are working to first-order in the
non-Gaussianity of the galaxy and PSF, we may simplify equation
(\ref{eq:oid1}) to give (for $pq>00$):

\beq
{c_{pq}\over c_{00}} \approx
\rme^{-\rmi m\chi}\left[
 \sum_{p'q'>00} Z^{p'q',00}_{pq}(\bar C_f,\bar D_f) \rme^{-\rmi
m'\alpha_f}
\beta^{(f)}_{p'q'}
+
\sum_{p''q''>00} Z^{00,p''q''}_{pq}(\bar C_f,\bar D_f) \rme^{-\rmi
m''\alpha_g}
\beta^{(g)}_{p''q''}
\right],
\label{eq:oid2}
\eeq
where we have worked to first order in the $\beta$'s, and used the result
that $Z^{00,00}_{00}=1$.

We next wish to relate the $c_{pq}$ to the $\beta^{(I)}_{pq}$.  This can
be done by use of the generators for point
transformations provided by BJ02.  To first order in the $\beta$'s and in
${\bf M}_I-{\bf M}_h$, and noting that
$\beta_{11}=\beta_{22}$, we get [see BJ02 eqs. (6-45) and (6-46)]:

\beq
\left\{\begin{array}{ll}
c_{11}/c_{00} \approx \mu & \\
c_{02}/c_{00} \approx {1\over 2\sqrt 2}\eta & \\
c_{20}/c_{00} \approx {1\over 2\sqrt 2}\eta^\ast & \\
c_{pq}/c_{00} \approx \beta^{(I)}_{pq} & pq\neq 11,02,20
\end{array}\right.,
\label{eq:hi}
\eeq
where $\mu$ is the dilation and $\eta$ is the shear required to transform
from a covariance ${\bf M}_I$ to ${\bf
M}_h$.  Explicitly:

\beq
{\bf M}_h = (1-2\mu){\bf M}_I - {\bf M}_I^{1/2}
\left( \begin{array}{cc} \Re\eta & \Im\eta \\ \Im\eta & -\Re\eta
\end{array} \right)
{\bf M}_I^{1/2},
\label{eq:hi2}
\eeq
where $\Re$ and $\Im$ denote real and imaginary parts, respectively.

Assuming that we know the PSF, $\beta^{(g)}_{p''q''}$ can be
determined.  We further know that
$\beta^{(f)}_{11}=\beta^{(f)}_{02}=\beta^{(f)}_{20}=0$.  Now truncate
equation (\ref{eq:oid2}) with a finite
set $L$ of Laguerre modes, i.e. we sum only over $p'q'\in L$.  We then
approximate the higher-order moments
$c_{pq}/c_{00}\approx \beta^{(I)}_{pq}$ for $pq\in\bar L\equiv
L\setminus\{00,11,02,20\}$ in accordance with equation
(\ref{eq:hi}).  This gives us a system of linear equations that can be
solved for the unknown $\beta^{(f)}_{pq}$'s:

\beq
\sum_{p'q'\in\bar L} Z^{p'q',00}_{pq}(\bar C_f,\bar D_f) \rme^{-\rmi
m'\alpha_f}
\beta^{(f)}_{p'q'} =
 \rme^{\rmi m\chi}\beta^{(I)}_{pq} - \sum_{p''q''\in\bar L}
Z^{00,p''q''}_{pq}(\bar C_f,\bar
D_f) \rme^{-\rmi m''\alpha_g}
\beta^{(g)}_{p''q''},
\label{eq:mat}
\eeq
for $pq\in\bar L$.  The $\beta$'s so determined can be substituted into
equation (\ref{eq:oid2}) to yield
$c_{11}/c_{00}$ and $c_{02}/c_{00}$, which then yield ${\bf M}_h$ via
equations (\ref{eq:hi}) and
(\ref{eq:hi2}).  Finally, we can determine ${\bf M}_f={\bf M}_h-{\bf
M}_g$ and then determine the intrinsic ellipticity
from equation (\ref{eq:edef}).

The above procedure is somewhat involved, and we illustrate it with a
simple example: that of $\bar L = \{22\}$.  That
is, we will correct for the radial fourth moment or kurtosis of the PSF
and intrinsic galaxy image.  Obviously in
ignoring the higher-order radial and angular moments we are
making an approximation, however the $b_{22}$ moment is present for
non-Gaussian galaxy profiles such as the exponential or de Vaucouleurs
profiles (and is positive in both
cases).  As we show in Section \ref{sec:sim}, the higher-order moments
can be important for some PSFs.
The angular moments $b_{pq}$ for $p\neq q$ are not present for pure
radial profiles, i.e. profiles whose isophotes are concentric ellipses
that share
principal axes and axis ratios.  If the angular moments, or the
higher-order radial moments $b_{33}$, $b_{44}$, etc. are important, these
would need to be included in
$\bar L$.

Because of the
shear and rotation invariance of the adaptive moments method, we can
assume without loss of generality that the PSF has
${\bf e}^{(g)}=0$, and that the measured object has $e^{(I)}_\times=0$
and $e^{(I)}=e^{(I)}_+\ge 0$, i.e. that it is
elongated along the $x$-axis.  (Alternatively, if a rounding kernel has
been applied to circularize the PSF, then we
already have
$e^{(g)}=0$, and no application of shear invariance is necessary.)  
 We then have:

\beq
{\bf M}_g=\left(\begin{array}{cc} {1\over 2}T_g & 0 \\ 0 & {1\over 2}T_g
\end{array}\right)
{\rm\hskip 0.1in and\hskip 0.1in}
{\bf M}_I=\left(\begin{array}{cc} {1\over 2}T_I(1+e^{(I)}) & 0 \\ 0 &
{1\over 2}T_I(1-e^{(I)}) \end{array}
\right) .
\label{eq:4start}
\eeq
Then to zeroeth order we determine:

\beq
{\bf M}_f|_{\rm 0th~order} = {\bf M}_I-{\bf M}_g
= \left( \begin{array}{cc} {1\over 2}[T_I(1+e^{(I)})-T_g] & 0 \\ 0 &
{1\over 2}[T_I(1-e^{(I)})-T_g] \end{array}\right)
.
\eeq
Following the procedure outlined in Appendix \ref{sec:conv}, we compute
the RDR decomposition of ${\bf M}_f^{1/2}
{\bf M}_I^{-1/2}$ and ${\bf M}_g^{1/2}{\bf M}_I^{-1/2}$; this gives
$\alpha_f=\alpha_g=\chi=0$, and:

\beqa
\bar C_g = 1-\bar C_f = T_g/[ T_I(1+e^{(I)})] \nonumber\\
\bar D_g = 1-\bar D_f = T_g/[ T_I(1-e^{(I)})]
\eeqa
Next we must construct the system of equations, (\ref{eq:mat}), which
requires knowledge of the $Z$ coefficients.  The
relevant $Z$ coefficients for the intrinsic image are:

\beqa
&& Z^{22,00}_{02} = Z^{22,00}_{20} = -{1\over 2\sqrt 2}\left[ {3\over
2}(\bar C_f^4-\bar D_f^4) - 2(\bar C_f^2-\bar
D_f^2)\right]
\nonumber\\
&& Z^{22,00}_{22} = {3\over 8}(\bar C_f^4+\bar D_f^4) + {1\over 4}\bar
C_f^2 \bar D_f^2
\nonumber\\
 && Z^{22,00}_{11} = -{3\over 4}(\bar C_f^4+\bar D_f^4) - {1\over 2}\bar
C_f^2 \bar D_f^2
   +\bar C_f^2 +\bar D_f^2
;
\eeqa
and those for the PSF are:

\beqa
&& Z^{00,22}_{02} = Z^{00,22}_{20} = -{1\over 2\sqrt 2}\left[ {3\over
2}(\bar C_g^4-\bar D_g^4) - 2(\bar C_g^2-\bar
D_g^2)\right]
\nonumber\\
&& Z^{00,22}_{22} = {3\over 8}(\bar C_g^4+\bar D_g^4) + {1\over 4}\bar
C_g^2 \bar D_g^2
\nonumber\\
&& Z^{00,22}_{11} = -{3\over 4}(\bar C_g^4+\bar D_g^4) - {1\over 2}\bar
C_g^2 \bar D_g^2
     +\bar C_g^2 +\bar D_g^2
.
\eeqa
From these we construct the linear system, equation (\ref{eq:mat}); in
this case it is only a $1\times 1$ system and
hence is easy to solve:

\beq
\beta^{(f)}_{22} = {1\over Z^{22,00}_{22}} \left[
\beta^{(I)}_{22} - Z^{00,22}_{22} \beta^{(g)}_{22}
\right].
\eeq
Then we can compute the correction ${\bf M}_I\mapsto{\bf M}_h$ due to
non-Gaussianity using equations
(\ref{eq:oid2}) and
(\ref{eq:hi}):

\beqa
\mu = c_{11}/c_{00} = Z^{22,00}_{11} \beta^{(f)}_{22} + Z^{00,22}_{11}
\beta^{(g)}_{22} \nonumber\\
\eta = 2\sqrt 2 c_{02}/c_{00} = 2\sqrt 2\left[Z^{22,00}_{02}
\beta^{(f)}_{22} + Z^{00,22}_{02} \beta^{(g)}_{22}
\right]
.
\label{eq:4end}
\eeqa
(Note that in this case, $\eta$ is real.  This is not necessarily the
case if we use $m\neq 0$ modes in our
basis $\bar L$.)  Finally, we compute ${\bf M})_h$ using equation
(\ref{eq:hi2}) and find that:

\beq
{\bf M}_f = {\bf M}_h - {\bf M}_g =
{1\over 2}
\left(\begin{array}{cc}
(1-2\mu-\eta)T_I(1+e^{(I)}) - T_g & 0 \\
0 & (1-2\mu+\eta)T_I(1-e^{(I)}) - T_g
\end{array}\right)
.
\eeq
The ellipticity ${\bf e}^{(f)}$ of the intrinsic galaxy image are easily
computable from ${\bf M}_f$ using the covariance-ellipticity relations
[equation (\ref{eq:edef})]:

\beqa
e^{(f)}_+ && =
{M^{(f)}_{xx}-M^{(f)}_{yy}\over M^{(f)}_{xx}+M^{(f)}_{yy}}
= \frac{
[(1-2\mu-\eta)T_I(1+e^{(I)}) - T_g] - [(1-2\mu+\eta)T_I(1-e^{(I)}) - T_g]
}{
[(1-2\mu-\eta)T_I(1+e^{(I)}) - T_g] + [(1-2\mu+\eta)T_I(1-e^{(I)}) - T_g]
}
\nonumber\\ &&
=
\frac{
(1-2\mu-\eta)(1+e^{(I)}) - (1-2\mu+\eta)(1-e^{(I)})
}{
(1-2\mu-\eta)(1+e^{(I)}) + (1-2\mu+\eta)(1-e^{(I)}) - 2T_g/T_I
}
=
\frac{
-\eta + (1-2\mu)e^{(I)}
}{
1-2\mu-\eta e^{(I)} - T_g/T_I
}
.
\label{eq:4corr}
\eeqa
This is equivalent to BJ02's method, equation (\ref{eq:bjr}), except that
the resolution factor is now:

\beq
R = \frac{
1-2\mu-\eta e^{(I)} - T_g/T_I
}{
-\eta + (1-2\mu)e^{(I)}
}e^{(I)}.
\label{eq:4corr2}
\eeq
If a shear has
been applied to circularize the PSF, the inverse shear must now be
applied to yield the physical ${\bf e}^{(f)}$.  The transformation of
ellipticities  ${\bf
e}\mapsto{\bf e}'$ under a shear $(\delta_+,\delta_\times)$ (i.e. a shear
that maps a circle to an ellipse of ellipticity $\delta$) is
given by [see, e.g. BJ02 eq. (2-13)]:

\beq
\left\{\begin{array}{l}
e'_+ = {1\over 1+\delta_+ e_+ +\delta_\times e_\times}\left[ e_+ +
\delta_+ + {\delta_\times\over\delta^2}(1-\sqrt{1-\delta^2})(e_\times
\delta_+ - e_+ \delta_\times)
\right]
\\
e'_\times = {1\over 1+\delta_+ e_+ +\delta_\times e_\times}\left[
e_\times + \delta_\times +
{\delta_+\over\delta^2}(1-\sqrt{1-\delta^2})(e_+ \delta_\times - e_\times
\delta_+)
\right]
\end{array}\right.,
\label{eq:composition}
\eeq
where $\delta^2 = \delta_+^2 + \delta_\times^2$; the transformation of
traces can be determined from conservation of area,
$T'\sqrt{1-e'^2}=T\sqrt{1-e^2}$.

[Warning: the ``shear composition'' operator ${\bf e}'=\delta\oplus{\bf
e}$ as defined by BJ02 and in equation (\ref{eq:composition}) does not
form a group because it is not associative.  This problem arises because
the shear ${\bf e}$ followed by the shear $\delta$ does not have the same
effect as the shear ${\bf e}'$ (they differ by an initial rotation).  For
this reason, it is probably better to think of $\oplus$ as representing
the transformation law for ellipticities under shear, rather than as a
composition operator.]

We can thus list the steps involved in PSF correction using the
moments $\bar L=\{22\}$.  If the PSF has been
circularized using a rounding kernel, the
steps marked with a $\star$ can be skipped.

\begin{list}{$\bullet$}{}
\item Compute the adaptive covariances ${\bf M}_I$ and ${\bf M}_g$ for
the galaxy and PSF, and the corresponding radial fourth moments
$\beta_{22}^{(I)}$ and
$\beta_{22}^{(g)}$.
\item $\star$ Apply a shear to circularize the PSF, i.e. use equation
(\ref{eq:composition}) with $\delta=-{\bf e}^{(g)}$.  The radial fourth
moments are unchanged by this
shear.
\item Apply a rotation so that the measured galaxy has $e^{(I)}_+>0$ and
$e^{(I)}_\times=0$.
\item Compute the dilation $\mu$ and shear $\eta$ in accordance with
equations (\ref{eq:4start}) through (\ref{eq:4end}).
\item Compute the intrinsic ellipticity of the galaxy in the sheared
rotated coordinate system using equation (\ref{eq:4corr}).
\item Undo the rotation.
\item $\star$ Shear to return the ellipticity to the unsheared unrotated
coordinate
system.  The shear is undone using equation
(\ref{eq:composition}) with $\delta={\bf e}^{(g)}$, i.e. the exact
opposite shear that was applied initially.
\end{list}

\section{The KSB95 Method in Laguerre Coefficients}
\label{sec:ksb}

Here we describe our implementation of the KSB95 shear estimator, updated
to include
the pre-seeing shear
polarizability \citep{1997ApJ...475...20L}.  For simplicity, we will
assume a Gaussian weight function.
We begin by computing the Laguerre expansion [see equation
(\ref{eq:lexpand})] of the measured object
$I$ and PSF $g$ to second order with a circular covariance matrix:

\beq
b^{(I)}_{pq} = 
\frac{1}{\sigma^2\sqrt{\pi\cdot p!q!}}
\int_{{\bf R}^2}
I({\bmath x}) \rme^{-(x^2+y^2)/2\sigma^2} \Lambda^\ast_{pq}
(x/\sigma,y/\sigma) \rmd^2{\bmath x}
,
\eeq
where $\sigma$ is the width of the weight function.  Then KSB's
ellipticity ${\bf e}^K$ (which in general is distinct from the adaptive
ellipticity!) is given by:

\beq
e^K_+ + \rmi e^K_\times = {\sqrt 2 b_{20}\over T_w} ,\; T_w =
b_{00}+b_{11}.
\eeq
Then the
shear polarizability hypertensor is given by:

\beq
[P^{\rm sh}]_{\alpha\beta} = [X^{\rm sh}]_{\alpha\beta} - [e^K]_\alpha
[e^{\rm sh}]_\beta,
\eeq
where:

\beq
[X^{\rm sh}]_{\alpha\beta} = 
\frac{b_{00} - b_{22}}{T_w} \delta_{\alpha\beta}
+
{\sqrt 6\over T_w }
\left(
\begin{array}{cc}
\Re b_{40} & \Im b_{40} \\
\Im b_{40} & -\Re b_{40}
\end{array}
\right)
\eeq
and

\beq
e^{\rm sh}_+ + \rmi e^{\rm sh}_\times = -\frac{ \sqrt 2 b_{20} + \sqrt 6
b_{31} }{T_w}
.
\eeq
The KSB/LK method also requires the smear polarizability hypertensor,
given by:

\beq
[P^{\rm sm}]_{\alpha\beta} = [X^{\rm sm}]_{\alpha\beta} - [e^K]_\alpha
[e^{\rm sm}]_\beta,
\eeq
where:

\beq
[X^{\rm sm}]_{\alpha\beta} =
\frac{b_{00} - 2b_{11} + b_{22}}{2T_w} \delta_{\alpha\beta}
+
{\sqrt 6\over 4T_w }
\left(
\begin{array}{cc}
\Re b_{40} & \Im b_{40} \\
\Im b_{40} & -\Re b_{40}
\end{array}
\right)
\eeq
and

\beq
e^{\rm sm}_+ + \rmi e^{\rm sm}_\times = -\frac{ \sqrt 2 b_{20} - \sqrt 6
b_{31} }{4T_w}.   
\eeq
In these equations, the $X$ contributions to the polarizabilities are due
to the effect of shear on the quadrupole moments $b_{20}$ whereas the
$e_\alpha e^{\rm
sh,sm}_\beta$ contributions are due to the effect of the shear on the
weighted trace $T_w$.  When the $X$ hypertensors are expressed in terms
of the Laguerre moments, the
decomposition into spin-0 (scalar) and spin-4
(traceless-symmetric) components is manifest.

\citet{1997ApJ...475...20L} then defines a pre-seeing shear
polarizability hypertensor to take account of the finite PSF
size:

\beq
[P^\gamma]_{\alpha\beta} = [P^{\rm sh}(I)]_{\alpha\beta} - [P^{\rm
sm}(I)]_{\alpha\gamma} [P^{\rm sh-1}(g)]_{\gamma\delta} [P^{\rm
sm}(g)]_{\delta\beta}.
\eeq
Our only remaining issue is the choice of the width $\sigma$ of the
weight function.  We use a weight function scale set by the requirement
$b_{11}=0$, i.e. we use an
adaptive circular variance.  This is appropriate, at least for the simple
case of a circular Gaussian object, for maximum significance detection
with a Gaussian filter of
variable width (see the discussion in Section
\ref{sec:multiscale}).  Following \citet{1998NewAR..42..137H}, we use the
same $\sigma^2$ for the PSF as for the object.

\end{document}